\documentclass[%
nofootinbib,
superscriptaddress,
 preprint,
showpacs,preprintnumbers,
 amsmath,amssymb,
 aps,
pra,
 longbibliography,
 lengthcheck,%
]{revtex4-1}

\usepackage[caption=false,parskip=100pt]{subfig}

\makeatletter\providecommand*{\diff}%
{\@ifnextchar^{\DIfF}{\DIfF^{}}}\def\DIfF^#1{%
\mathop{\mathrm{\mathstrut d}}%
\nolimits^{#1}\gobblespace}\def\gobblespace{%
\futurelet\diffarg\opspace}\def\opspace{%
\let\DiffSpace\!%
\ifx\diffarg(%
\let\DiffSpace\relax\else\ifx\diffarg[%
\let\DiffSpace\relax\else\ifx\diffarg\{%
\let\DiffSpace\relax\fi\fi\fi\DiffSpace}
\makeatother

\usepackage{amsthm}
\newtheorem{lemma}{Lemma}  
\newtheorem{theorem}{No-go theorem}
\newtheorem{corollary}{Corollary}[theorem]
\newenvironment{remark}[1][Remark.]{\begin{trivlist}
\item[\hskip \labelsep {\bfseries #1}]}{\end{trivlist}}

\usepackage[mathcal]{euscript}
\usepackage[greek,english]{babel}
\usepackage{cases}
\usepackage{enumerate}
\usepackage{xspace}
\usepackage{type1cm}
\usepackage{longtable}
\usepackage{graphicx}
\usepackage{epstopdf}
\DeclareGraphicsExtensions{.eps,.png}
\usepackage{dcolumn}
\usepackage{bm}
\usepackage{hyperref}
\usepackage{amsthm}
\usepackage{ifthen, paralist}
\usepackage{accents}
\usepackage[cal=boondoxo]{mathalfa} 
\usepackage{arcs}
\usepackage{color}

\newcommand{\bra}[1]{\mathop{\langle{#1}|}\nolimits}
\newcommand{\ket}[1]{\mathop{|{#1}\rangle}\nolimits}
\newcommand{\midop}[1]{\mathop{\langle #1 \rangle}\nolimits}
\newcommand{\FREE}{\renewcommand{\bra}{}\renewcommand{\ket}{}}
\newcommand{\matel}[3]{{\FREE\mathop{\langle{#1}|#2\,|{#3}\rangle}\nolimits}}
\newcommand{\scpr}[2]{{\FREE\mathop{\langle{#1}|{#2}\rangle}\nolimits}}

\newcommand{\LR}[0]{{\mathop{\leftrightarrow}}}

\newcommand{\<}[1]{}
\newcommand{\pder}[2]{\mathop{\frac{\partial #1}{\partial #2}}}
\newcommand{\tpder}[2]{\mathop{\tfrac{\partial #1}{\partial #2}}}
\newcommand{\tder}[2]{\mathop{\tfrac{d #1}{d #2}}}
\newcommand{\der}[2]{\mathop{\frac{d #1}{d #2}}}
\newcommand{\idx}[1]{\mbox{\normalfont\scriptsize #1}}

\let\originalleft\left
\let\originalright\right
\renewcommand{\left}{\mathopen{}\mathclose\bgroup\originalleft}
\renewcommand{\right}{\aftergroup\egroup\originalright}

\usepackage{xifthen}
\newcommand{\proj}[2][]{\FREE
\ifthenelse{\equal{#1}{}}{\mathop{|{#2}\rangle\langle{#2}|}}{\mathop{|{#1}\rangle\langle{#2}|}}%
\nolimits}

\newcommand{\ehat}[1]{\expandafter\hat#1}

\newcommand{\Tr}{\mathop{\rm{Tr}}}
\newcommand{\es}[1]{{\mathop{\mathtt{#1}}}} 

\newcommand{\Lrel}{{\cal L}_{\idx{rel}}}
\newcommand{\Lbd}{{\cal L}^{\idx{lbd}}}
\newcommand{\Bdn}{{\cal B}}
\newcommand{\Sgn}{\mathop{\mathrm{sign}}}
\newcommand{\defeq}{\mathop{\stackrel{\idx{def}}{=}}}
\DeclareMathOperator*{\argmin}{arg\,min}

\let\oldgather = \gather
\let\endoldgather = \endgather
\renewenvironment{gather}[0]{\par\nobreak\noindent\oldgather}{\endoldgather}
\let\oldalign = \align
\let\endoldalign = \endalign
\renewenvironment{align}[0]{\par\nobreak\noindent\oldalign}{\endoldalign}

\newcommand*{\appref}[1]{
\def\temp{#1}\ifx\temp\empty{\leavevmode\unskip supplemental material}\ignorespaces
\else
    \leavevmode\unskip Appendix~\ref{#1}\ignorespaces
\fi
}

\newcommand{\APPBIBITEM}{
}

\newcommand{\LvnG}{{\cal L}^{\idx{G}}}
\newcommand{\ffG}{\tilde{{f}}^{G}}
\newcommand{\AG}{A^{\idx{G}}}
\newcommand{\kkappaG}{\boldsymbol{\mu}}

\begin{document}
\title{Quantum friction: environment engineering perspectives}

\author{Dmitry V. Zhdanov}
\email{dm.zhdanov@gmail.com}
\affiliation{Department of Chemistry, Northwestern University, 2145 Sheridan Road, Evanston, Illinois 60208-33113 USA}
\author{Denys I. Bondar}
\affiliation{Princeton University, Princeton, NJ 08544, USA}
\author{Tamar Seideman}
\email{t-seideman@northwestern.edu}
\affiliation{Department of Chemistry, Northwestern University, 2145 Sheridan Road, Evanston, Illinois 60208-33113 USA}

\begin{abstract}
We prove a generalization of the Lindblad's fundamental no-go result: A quantum system cannot be completely frozen and, in some cases, even thermalized via translationally invariant dissipation -- the quantum friction. Nevertheless, a practical methodology is proposed for engineering nearly perfect quantum analogs of classical friction within the Doppler cooling framework. These findings pave the way for hallmark dissipative engineering (e.g.\ nonreciprocal couplings) with atoms and molecules.
\end{abstract}
\maketitle
\section{Introduction} Heat dissipation in a broad variety of phenomena, ranging from spontaneous emission to chemical dynamics in solvents, is attributed to friction forces that can be velocity-dependent but are coordinate-invariant. Description of such dynamics has meet serious challenges since the dawn of quantum theory of open systems \cite{BOOK-Razavy}. As early as in 1976, around the time of publishing the equation now bearing his name \cite{1976-Lindblad}, Goran Lindblad also came up with a counterintuitive no-go result when attempting to quantize a classical damped harmonic oscillator \cite{1976-Lindblad-a}. He showed that the quantum analog of the classical velocity-dependent friction (i.e.,\ translationally invariant quantum Markovian dissipative process) is unable to equilibrate an oscillator to any temperature $\theta{=}k_{\idx B}T$, including $\theta{=}0$ \footnote{It turned out that the 
stochastic fluctuations accompany quantum friction in all cases complying with the detailed balance condition with no extra assumptions (thermalized reservoir, linear response etc.) needed for standard fluctuation-dissipation theorems.}.

Lindblad's reasoning is applicable to multidimensional harmonic oscillators \cite{1985-Dodonov} and specific dissipators\footnote{Specifically, when all $A_k$ in Eqs.~\eqref{Quantum_Liouville_equation_nD}, \eqref{theorem:trans_inv_Lbd} are linear in $\hat x$ and $\hat p$.} only. Nevertheless, his no-go finding 
has long been believed to hold universally and has been accepted
without a proof \cite{1997-Kohen}. Here we present such a proof for an arbitrary quantum system in the case $\theta{=}0$ and also for a harmonic oscillator without any restriction on $\theta$. This result calls for revisiting the applicability and consistency of quantum friction models. In this Letter, we focus on implications for quantum reservoir engineering (QRE), an emerging method for controlling dynamical processes by dissipative environmental interactions. 
Both experiments \cite{2011-Krauter} and theory \cite{2012-Muller,2014-Kronwald,2012-Eremeev,2010-Pielawa,2012-Koga,2012-Marcos,2009-Verstraete,2012-Murch,2006-Pechen,2006-Pechen-2}  
provide evidence that QRE offers a viable alternative to
the traditional methods of 
coherent control (e.g.,\ it provides a self-suf\-fi\-cing  framework for quantum information processing \cite{2013-Kastoryano,2011-Barreiro}). QRE also features unique capabilities on rendering the quantum interactions directional \cite{2015-Metelmann} and cancelling the noise by ``no-knowledge'' measurements \cite{2014-Szigeti}. 
Recent studies of QRE for an ensembles of cold Rydberg atoms revealed plethora of novel dissipation-assisted phenomena including bond formation \cite{2013-Lemeshko}, quantum phase transitions \cite{2014-Lee}, exciton transport \cite{2015-Schempp,2015-Schonleber}, and energy redistribution \cite{2014-Everest,2014-Lesanovsky}. Promising results were also obtained for other systems including optomechanical arrays \cite{2012-Tomadin,2014-Woolley} and transmons \cite{2013-Shankar,2014-Cohen,2013-Leghtas}. The mentioned works mostly use optical or microwave cavities as reservoirs offering highly tunable relaxation by adjusting nonlinear couplings and photon loses. However, this type of reservoirs requires bulky, intricate, and costly equipment preventing a broad variety of applications in physics and chemistry from taking advantage of QRE. 

In this Letter, we argue that quantum friction is a promising and powerful resource for pushing the boundaries of QRE. First, it is demonstrated that the fundamental decoherence limits imposed by the no-go constraints do not preclude the effective quantum state engineering including nearly perfect cooling. Second, we show how the established quantum-optics and cavity-electrodynamics technologies can be readily utilized to prototype customizable quantum friction forces in the laboratory. 
Third, the utility of these forces for designing QRE-based gadgets is illustrated by proposing a mechanical analog of a nonreciprocal photonic device \cite{2015-Metelmann}.

The Letter is set up as follows. We begin with formalizing the notion of quantum friction following Refs. \cite{1976-Lindblad-a,2016-Bondar} and then present rigorous but yet physical formulations of our key no-go theorems deferring details to the \appref{}. The physical meaning of these theorems is further clarified by addressing the above listed three arguments for advocating the quantum friction candidacy for QRE. 


\section{The formal definitions of quantum friction and no-go theorems\label{@SEC:gen_form}}
The starting point for our analysis is the von-Neumann equation for density matrix $\hat\rho$ of a Markovian open quantum system with $N$ degrees of freedom
\newcommand{\pp}{\boldsymbol{p}}
\newcommand{\xx}{\boldsymbol{x}}
\newcommand{\ff}{\boldsymbol{f}}
\newcommand{\fc}{\tilde f}
\newcommand{\fabs}{f}
\newcommand{\argf}{\varphi}
\newcommand{\fciso}{\tilde f^{\idx{iso}}}

\newcommand{\ggamma}{\boldsymbol{\gamma}}
\newcommand{\kkappa}{\boldsymbol{\kappa}}
\newcommand{\dxx}{\boldsymbol{\delta x}}
\newcommand{\set}[1]{\boldsymbol{#1}}
\newcommand{\rhoth}[1]{\hat\rho^{\idx{th}}_{#1}}
\newcommand{\frnforce}{F^{\idx{fr}}}
\newcommand{\ffrnforce}{\boldsymbol{F}^{\idx{fr}}}
\begin{subequations}
\label{Quantum_Liouville_equation_nD}
\begin{gather}\label{Quantum_Liouville_equation_nD(a)}
\pder{}{t}\hat\rho{=}{\cal L}[\hat\rho]{=}{-}\frac{i}{\hbar}[\hat H,\hat\rho]{+}\Lrel[\hat\rho],\\
\hat H{=}H(\hat\pp,\hat\xx){=}\sum_{n=1}^N\frac{\hat p_n^2}{2 m_n}{+}U(\hat\xx).\label{Quantum_Hamiltonian_nD}
\end{gather}
\end{subequations}
Here $\hat\xx{=}\{\hat x_1,...,\hat x_N\}$ and $\hat\pp{=}\{\hat p_1,...,\hat p_N\}$ are vectors of the canonical operators of positions and momenta, respectively, $\hat H$ is the system Hamiltonian, and $\Lrel$ is the dissipative superoperator accounting for system-environment interactions. 

Lindblad \cite{1976-Lindblad} have shown that Eq.~\eqref{Quantum_Liouville_equation_nD(a)} retains the physical meaning for all feasible states $\hat\rho$ only if $\cal{L}_{\idx{rel}}$ has the structure
$
\Lrel[\hat\rho]{=}\sum_k\Lbd_{\hat L_k}[\hat\rho], 
$ 
 where
\begin{gather}\label{_Lindbladian_definition}
\forall \hat L:\Lbd_{\hat L}{\defeq}\hat L{\odot}\hat L^{\dagger}{-}\frac12(\hat L^{\dagger}\hat L{\odot}{+}{\odot}L^{\dagger}\hat L).
\end{gather}
Throughout the Letter, we will focus on the case where the Ehrenfest relations for average positions and momenta resemble the Newtonian motion in a potential $U(\xx)$ damped by friction forces $\ffrnforce$
\begin{subequations}
\label{ODM-eqs'}
\begin{gather}
\label{ODM-dp/dt'}
\tder{}{t}\midop{\hat p_n}{=}{-}\midop{\tpder{}{\hat x_n}U(\hat \xx)}{+}\midop{\ehat\frnforce_n},\\
\label{ODM-dx/dt'}
\tder{}{t}\midop{\hat x_n}{=}\tfrac{1}{m}\midop{\hat p_n}.
\end{gather}
\end{subequations}
The classical friction forces are normally treated as position-independent functions of velocities: $\ffrnforce{=}\ffrnforce(\pp)$. In order to preserve this property in the quantum case, we must require $\cal{L}_{\idx{rel}}$ to be translationally invariant, i.e.,
\begin{gather}\label{Quantum_Liouville_trans_inv_nD}
\forall \hat\rho: [\hat\pp,\Lrel[\hat\rho]]{=}\Lrel[[\hat\pp,\hat\rho]].
\end{gather}

The following lemma (originally formulated by B. Vacchini \cite{2005-Petruccione,2005-Vacchini,2009-Vacchini} based on theorems by A. Holevo \cite{1995-Holevo,1996-Holevo}) provides the generic form of operators $\hat L$ in Eq.~\eqref{_Lindbladian_definition} consistent with the criterion \eqref{Quantum_Liouville_trans_inv_nD}.
 
\begin{lemma}[The proof is in \appref{@APP:theo:trans_inv_Lbd}]\label{@theo:trans_inv_Lbd}
Any translationally invariant superoperator $\Lrel$ of the Lindblad form (see Eq.~\eqref{_Lindbladian_definition}) can be represented as%
\footnote
{
The Gaussian dissipators $\Lbd_{\kkappaG_k\hat\xx{+}\ffG_k(\hat\pp)}$ $(\kkappaG_k{\in}\mathbb{R}^N)$ hereafter are 
treated as limiting case of Eq.~\eqref{theorem:trans_inv_Lbd} with $\kkappa_k{\to}0$ (see  \appref{@APP:theo:trans_inv_Lbd}).
}
\begin{gather}\label{theorem:trans_inv_Lbd}
\Lrel{=}\sum_k\Lbd_{\hat A_k} \mbox{ with } \hat A_k
{\defeq}e^{{-}i\kkappa_k\hat\xx}\tilde f_k(\hat\pp),
\end{gather}
where $\kkappa_k$  are $N$-dimensional real vectors, and $\tilde f_k$ are complex-valued functions. The converse holds as well.
\end{lemma}

In the case of an isotropic environment, the terms in Eq.~\eqref{theorem:trans_inv_Lbd} additionally must appear in pairs, so that $\Lrel{=}\sum_k\Bdn_{\kkappa_k,\fciso_{k}}$, where
\begin{gather}\label{theorem:trans_inv_Lbd-isotropic}
\Bdn_{\kkappa_k,\fciso_{k}}{\defeq}\Lbd_{\hat A_k^+}{+}\Lbd_{\hat A_k^-} \mbox{ with }\hat A_k^{\pm}{=}e^{{{\mp}}i\kkappa_k\hat\xx}\fciso_k({\pm}\hat\pp).
\end{gather}

Substitution of \eqref{theorem:trans_inv_Lbd} into Eq.~\eqref{Quantum_Liouville_equation_nD} allows to explicitly find the classical analog of quantum friction force in Eq.~\eqref{ODM-dp/dt'}
\begin{gather}\label{_Bondarian_gen}
\ffrnforce(\pp){=}{-}\sum_k\hbar\kkappa_{k}|\tilde f_k(\pp)|^2{=}{-}\sum_k\hspace{-3pt}\sum_{\alpha{=}{\pm}1}\hspace{-5pt}\alpha\hbar\kkappa_{k}|\fciso_k(\alpha\pp)|^2.
\end{gather}
With this, quantum and classical frictions are fundamentally different beasts. In the classical case the energy dissipation cools the system to complete rest in accordance with the second law of thermodynamics. However, this is never the case for quantum friction: 

\begin{theorem}[The proof is in \appref{@APP:theo:no_go(T=0)}] \label{@theo:no_go(T=0)}
No translationally invariant Markovian process of form \eqref{Quantum_Liouville_equation_nD} and \eqref{theorem:trans_inv_Lbd} with non-(quasi)periodic potential $U(\xx)$ can steer the system to any eigenstate of $\hat H$, including the ground state.
\end{theorem}

\newcommand{\BFn}{B}
\newcommand{\llambda}{\boldsymbol{\lambda}}
\newcommand{\zzero}{\boldsymbol{0}}

The above result can be strengthened for a special class of quantum systems. Let $\BFn(\pp,\llambda)$ denote the Blokhintsev function, which is related to Wigner quasiprobability distribution $W(\pp,\xx)$ as
\begin{gather}\label{_Blokhintsev_function}
\BFn(\pp,\llambda){=}\int_{{-}\infty}^{\infty}\ldots\int_{{-}\infty}^{\infty} e^{i \llambda\xx} W(\pp,\xx)\diff^N\xx.
\end{gather}

\begin{theorem}[The proof is in \appref{@APP:theo:no_go(T>0)}]\label{@theo:no_go(T>0)} 
Suppose that the Blokhintsev function $\BFn_{\theta}(\pp,\llambda)$ of the thermal state $ \rhoth{\theta}{\propto}e^{-\frac{\hat H}{\theta}}$ characterized by temperature $k_{\idx{B}}T{=}\theta$ is such that
\begin{subequations}\label{_B(p,lambda)-features}
\begin{gather}\label{_B(p,lambda)-features-a'}
\forall\pp,\llambda: \BFn_{\theta}(\pp,\llambda){>}0,~~\BFn_{\theta}(\pp,{-}\llambda){=}\BFn_{\theta}(\pp,\llambda),\\
%
\forall\pp{\ne}\zzero,\llambda{\ne}\zzero: \BFn_{\theta}(\pp,\llambda){<}\BFn_{\theta}(\zzero,\zzero).\label{_B(p,lambda)-features-b'}
\end{gather}
\end{subequations}
Then, no translationally invariant Markovian process of form \eqref{Quantum_Liouville_equation_nD} and \eqref{theorem:trans_inv_Lbd} can asymptotically steer the system to $\rhoth{\theta}$.
\end{theorem}

Using Eq.~\eqref{_Blokhintsev_function} and the familiar formula for the thermal state Wigner function \cite{1949-Bartlett}, it is easy to check that the criteria \eqref{_B(p,lambda)-features} are satisfied for any $\theta$ in the case of a quadratic potential $U$.

\begin{corollary}\label{@cor:no_go_qho(T>0)}
No translationally invariant Markovian process of form \eqref{Quantum_Liouville_equation_nD} and \eqref{theorem:trans_inv_Lbd} can steer the quantum harmonic oscillator into a thermal state of form $ \rhoth{\theta}{\propto}e^{{-}\frac{\hat H}{\theta}}$.
\end{corollary}

\section{Physical meaning of the no-go theorems\label{@SEC:implications}}
The genesis of the no-go results can be traced on the example of isotropic friction \eqref{theorem:trans_inv_Lbd-isotropic} in the limit when $\forall k: (\hbar\kkappa_k)^2{\ll}\midop{\hat\pp^2}$. In the free-particle case $U(\xx){=}0$, the expectation value of any observable of the form $h(\hat\pp)$ evolves as
\begin{gather}\notag
\tder{}{t}\midop{h(\hat\pp)}
{\simeq}\sum_{n{=}1}^N\midop{\frnforce_n(\hat\pp)\tpder{h(\hat\pp)}{\hat p_n}{+}\sum_{l{=}1}^ND_{l,n}(\hat\pp)\tpder{^2h(\hat\pp)}{\hat p_{l}\partial\hat p_{n}}},
\end{gather}
where $D_{l,n}(\hat\pp){=}\tfrac{\hbar^2}2\sum_k\sum_{\alpha{=}{\pm}1}\fciso_k(\alpha\hat\pp)^2\kappa_{l,k}\kappa_{n,k}$. 
This implies that the momentum probability distribution $\varpi(\pp){=}\Tr[\delta(\pp{-}\hat\pp)\hat\rho]$ satisfies the 
Fokker-Planck equation
\begin{gather}\label{_Fokker_Planck_equation}
\der{\varpi(\pp)}{t}{=}\sum_{n}\left({-}\pder{\frnforce_n(\pp)\varpi(\pp)}{p_n}{+}\sum_l\pder{^2D_{n,l}(\pp)\varpi(\pp)}{p_n\partial p_l}\right).
\end{gather}
The diffusion terms in Eq.~\eqref{_Fokker_Planck_equation} manifest the inherent presence of stochastic quantum fluctuations accompanying the energy dissipation even at $T{=}0$. Hence, the no-go theorems \ref{@theo:no_go(T=0)} and \ref{@theo:no_go(T>0)} can be viewed as generalized quantum fluctuation-dissipation theorems. 

The origin of quantum fluctuations can be rationalized as follows. In classical mechanics a friction terms as in Eq.~\eqref{ODM-dp/dt'} only decelerate the particles leaving their instant spatial distribution, and hence the potential energy $\midop{U}$, intact. This is not the case in quantum mechanics where the operators $\hat p_n$ and $\hat x_n$ are coupled through the canonical commutation relation, as can be seen from the friction-induced changes in the
second-order moments:
\newcommand{\dertH}{{\mathfrak{d}}_t}
\begin{subequations}\label{_2-nd_moments}
\begin{align}\label{_2-nd_moments_p^2}
\midop{\Lrel^{\intercal}[\hat p_n^2]}&{=}{-}\hbar\sum_k2{\kappa_{k,n}}\fabs_k(\hat\pp)^2(\hat p_n{-}\tfrac{\hbar\kappa_{k,n}}2),\\
\label{_2-nd_moments_x^2}
\midop{\Lrel^{\intercal}[\hat x_n^2]}&{=}{-}\hbar\sum_k\midop{\left\{\hat x_k,\fabs_k(\hat\pp)^2\tpder{\argf_k(\hat\pp)}{\hat p_n}\right\}_{+}}{+}\notag\\
&\hbar^2\sum_k\midop{\left(\tpder{\fabs_k(\hat\pp)}{\hat p_n}\right)^2{+}\left(\tpder{\argf_k(\hat\pp)}{\hat p_n}\right)^2\fabs_k(\hat\pp)^2},
\end{align}
\end{subequations}
where 
$\fabs_k(\pp){=}|\fc_k(\pp)|$, $\argf_k(\pp){=}\arg(\fc_k(\pp))$, and the symbol $^{\intercal}$ denotes the superoperator transpose. 
The first summation in Eq.~\eqref{_2-nd_moments_x^2} cancels out for any thermal state $ \rhoth{\theta}$, so that $\forall\theta,\fc_k: \midop{\Lrel^{\intercal}[\hat x_n^2]}{>}0$. In particular, the quantum friction always increases the potential energy in the harmonic oscillator case $U(\xx){=}\sum_{n{=}1}^N\frac{m\omega_n^2}2x_n^2$. This provides the physical rationale behind corollary~\ref{@cor:no_go_qho(T>0)}.

It is instructive to consider the case of a one-dimensional quantum harmonic oscillator in detail and to explore the extent at which one can beat the no-go results by intelligent reservoir engineering. From this point, we will omit everywhere the dimension subscript $n$. Our goal is to minimize the Bures distance $D_{\idx{B}}(\left.\hat\rho\right|_{t\to{\infty}},\rhoth{\theta_{0}})$ between the equilibrium state $\left.\hat\rho\right|_{t\to{\infty}}$ and the thermal state $\rhoth{\theta_{0}}$ for a given temperature $\theta_0$. A reasonable strategy is to search for a quantum friction term whose action does not change the energy distribution of the thermal state $\rhoth{\theta}$, i.e.,
\begin{gather}\label{_energy_distribution_conservation}
\forall \alpha: \Tr[\rhoth{\theta_0}\Lrel^{\intercal}[e^{-\alpha\hat H}]]{=}0.
\end{gather}
In the case $\theta_0{=}0$ the objective \eqref{_energy_distribution_conservation} can be reformulated as the variational problem for the parameters of Eq.~\eqref{theorem:trans_inv_Lbd}:
\begin{gather}\label{_energy_distribution_conservation(T=0)}
\{\kappa_k^{\idx{opt}},\tilde f_{k}^{\idx{opt}}\}{=}\argmin_{\{\kappa_k,\tilde f_{k}\}}\Tr[\rhoth{0}\Lrel^{\intercal}[\hat H]].
\end{gather}
The solution of the problems \eqref{_energy_distribution_conservation} and \eqref{_energy_distribution_conservation(T=0)} is
\begin{gather}\label{_optimal_f_k}
\tilde f_k^{\pm}(p){=}c_k e^{p \beta \hbar  \lambda_k^{\pm}},~~\lambda_k^{\pm}{=}\frac{\kappa_k}{r(\theta_0)}\left(1{\pm} \sqrt{1{-}r^2(\theta_0)}\right),
\end{gather} 
where $c_k$ and $\kappa_k$ are arbitrary real constants, $r(\theta){=}\tanh(\frac{\hbar\omega}{2 \theta }){=}\hbar\omega/(2\Tr[\hat H\rhoth{\theta}])$, and $\beta{=}(m\hbar\omega)^{-1}$. Both the $\pm$ branches are expected to have stable stationary points near $\rhoth{\theta_0}$ since $\Sgn(\Tr[\hat H\Lbd_{\hat A_k}[\rhoth{\theta}]]){=}\Sgn(\theta_0{-}\theta)$. Indeed, $\Tr[\hat H\Lbd_{\hat A_k}[\rhoth{\theta}]]{=}\frac{c_k^2}{\omega}\gamma^{\idx{en}}_k(\theta)(r(\theta){-}r(\theta_0))\Tr[\hat H\rhoth{\theta}]$, where
\begin{gather}
\gamma^{\idx{en}}_k(\theta){=}2\omega{\beta\hbar^2\kappa_k \lambda_k e^{\frac{\beta\hbar^2\lambda_k^2}{r(\theta)}}}{>}0.
\end{gather}

\begin{figure}[h!]
\centering\includegraphics[width=0.6\columnwidth]
{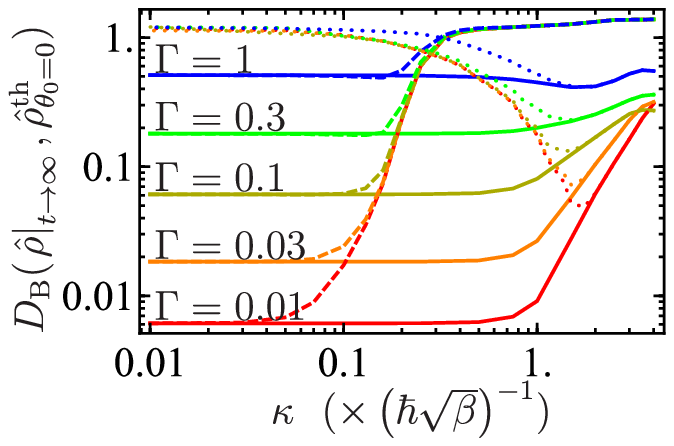}
\caption{The Bures measure of the quality of thermalization of a harmonic oscillator to the ground state by an isotropic quantum friction process $\Lrel{=}\Bdn_{\kappa,\fciso}$ with $\fciso(p)$ defined by Eq.~\eqref{_optimal_f_k} and $c{=}\sqrt{\Gamma}\omega/\sqrt{\gamma^{\idx{en}}(0)}$ as a function of $\kappa$ and $\Gamma$ (solid lines). The dotted lines represent the performance of the clipped versions of $\fciso(p)$. Also shown is the case when the function $\fciso(p)$ is approximated as $g(p){=}\tilde c_{1}\frac{(p-\tilde c_{3)}}{\tilde c_{2}^2{+}(p-c_{3})^2}$ of form \eqref{_coherent_Doppler_f(p)} with real parameters $\tilde c_{i}$ chosen such that $\left.\pder{^l}{p^l}(\tilde f(p){-}g(p))\right|_{p=0}{=}0$ for $l{=}0,1,2$ (dashed lines).\label{@FIG.01'}}
\end{figure}

A characteristic feature of the optimal functions $\tilde f_k^{\pm}(p)$ is extended exponential tails in the region $\kappa_kp{<}0$. These tails are counterintuitive because they increase the kinetic energy of the oscillator (see Eqs.~\eqref{_Bondarian_gen} and \eqref{_2-nd_moments_p^2}). However, according to Eq.~\eqref{_2-nd_moments_x^2} the energy change of the quantum oscillator additionally depends on the slopes of $\tilde f_k^{\pm}(p)$. Because of this quantum mechanism, clipping the ``endothermic'' tails in the region $\kappa_kp{<}0$ increases the net heating. This effect is illustrated in Fig.~\ref{@FIG.01'} depicting the results of numerical analysis of solutions \eqref{_optimal_f_k}. Figure~\ref{@FIG.01'} shows that a high-quality thermalization is readily achieved via tuning the free parameters $c_k$ and $\kappa_k$. Additionally, the quality weakly depends on the profile of $\tilde f_k(p)$ outside the region $|p|{<}\beta^{{-}\frac12}$. This offers freedom to replace the exact exponentially diverging solutions \eqref{_optimal_f_k} with physically feasible approximations (an example is shown by dashed lines). 

\section{Quantum friction in the laboratory\label{@SEC:phys_meaning}}
As a heuristic argument, note that Eq.~\eqref{_Fokker_Planck_equation} is similar to the Fokker-Planck equation describing the dynamics of atoms in laser fields undergoing Doppler cooling (see, e.g., Ref.~\cite{1992-Berg-Sorenson}). The interpretation of Doppler cooling as a quantum friction phenomenon is justified in \appref{@APP:phys_meaning} (see also Ref.~\cite{1996-Poyatos}). Here we provide a brief summary of the argument.

\begin{figure}[t!]
\centering
  \subfloat[]
     {\includegraphics[width=0.2\textwidth]{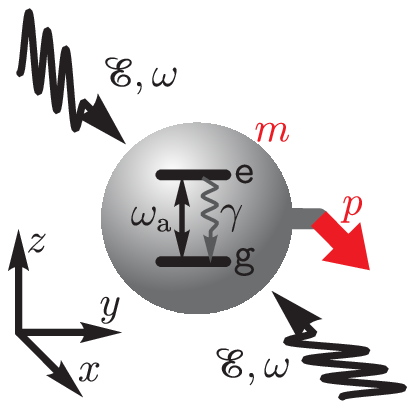}\label{@FIG.02'}}\hspace{15pt}
  \subfloat[]
     {\includegraphics[width=0.2\textwidth]{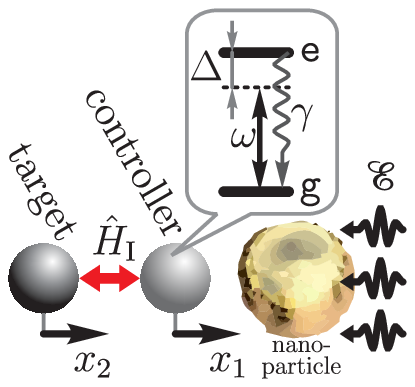}\label{@FIG.02b}}
\caption{
(a) The Doppler cooling model.
(b) ``Sympathetic'' nonreciprocal coupling scheme.
}
\end{figure}

Consider the scheme shown in Fig.~\ref{@FIG.02'}, where an atom is subject to
two orthogonally polarized counterpropagating beams of the same field amplitude $\cal E$ and carrier frequency $\omega$. 
We assume that $\omega$ is close to the frequency $\omega_{\idx{a}}$ of the transition $\es{g}\LR\es{e}$ between the ground $\es g$ and degenerate excited $\es e$ electron states of $s$- and $p$-symmetries, respectively. Let $d$ be the absolute value of the transition dipole moment and $\gamma$ be the excited state spontaneous decay rate. For simplicity, we consider only the case when the non-radiative decay mechanism is dominant and $\xi{=}\frac12{\cal E}d{\ll}\Delta{=}\omega_{\idx{a}}{-}\omega$ (the weak field limit). Then, the translational motion of atom can be fully characterized using Eq.~\eqref{Quantum_Liouville_equation_nD} with the isotropic friction term $\Lrel{=}\Bdn_{\kappa,\fciso}$. The specific form of $\fciso$ depends on the radiation coherence time $t_{\idx{coh}}$. In the case of a coherent CW laser with $t_{\idx{coh}}{\gg}\Delta^{-1},\gamma^{-1}$ one has
\begin{gather}\label{_coherent_Doppler_f(p)}
\fciso(p){=}\tfrac{|\xi|}{\hbar}\sqrt{\gamma}{\Delta_1(p)}\left({(\tfrac{\gamma}2)^2{+}\Delta_1(p)^2}\right)^{-1},
\end{gather}
where $\Delta_1(p){=}\Delta{-}\frac{\kappa( p{+}\frac{\hbar\kappa}{2})}{m}$ and $m$ is the atomic mass.
In the opposite limit of incoherent illumination with $\gamma^{-1}{\gg}t_{\idx{coh}}{\gg}\Delta_1^{-1}(p)$, the shape of $\tilde f(p)$ is defined by the radiation spectral density $I(\omega)$ of the beams:
\begin{gather}
\fciso(p){=}{\pi d}{\hbar}^{-1}\sqrt{I(\omega{+}\Delta_1(-p))/(2 c)}.
\end{gather}
This physical interpretation of quantum friction clarifies the deterioration of quality of cooling with increasing $|\kappa|$ observed in Fig.~\ref{@FIG.01'}. The analysis presented in \appref{@APP:phys_meaning} shows that $|\hbar\kappa|$ is the characteristic change of atomic momentum after a random photon absorption, whereas $|\fciso(p)^2|$ defines the absorption rate. In the case of small $|\kappa|$ and large $|\fciso(p)^2|$ the effect of an individual photon absorption on the translational motion is negligible, and the optical impact can be described in terms of the net radiational pressure whose fluctuations are negligible due to averaging over a large number of events. The opposite case of large $|\kappa|$ and small $|\fciso(p)^2|$ corresponds to the strong shot noise limit when the stochastic character of the absorption is no longer moderated by massive averaging. Note that a similar interpretation applies to quantum statistical forces in Ref.~\cite{2016-Vuglar}. 


\section{Nonreciprocal couplings\label{@SEC:nonreciprocal}} 
Let us finally demonstrate that the quantum friction can be used for non-trivial quantum engineering. Nonreciprocal optical and optomechanical devices have gained attention since they enable novel signal processing and quantum control applications. One says that two quantum systems (``controller'' and ``target'') are coupled nonreciprocally if the target's dynamics depends on the controller's state, whereas the reverse quantum feedback is absent. 
Such a one-way coupling is solely an open system phenomenon and not possible if the controller and target form together a closed system. It was recently shown that the nonreciprocity can be implemented in coupled dissipative optical cavities \cite{2015-Metelmann}. However, this proposal is not extendible to mechanical systems because the required interactions cannot be engineered. For instance, certain parameters in Eqs. (6) and (11) of Ref.~\cite{2015-Metelmann} cannot be made complex valued. Here we overcome this obstacle by means of quantum friction.

\newcommand{\nrcA}{\hat B}
\newcommand{\nrcf}{\tilde g}
\newcommand{\nrcF}{F^{\idx{nr}}}
\newcommand{\nrcD}{D^{\idx{nr}}}
Consider two coupled 
oscillators with the Hamiltonian
\begin{gather}\nonumber
\hat H{=}\sum_{n=1}^2\hat H_n{+}\hat H_{\idx I},~\hat H_n{=}\frac{\hat p^2_n}{2 m_n}{+}
U_n(\hat x_n)
,~\hat H_{\idx I}{=}\hat x_1\chi(\hat x_2),
\end{gather}
where the function $\chi(x)$ specifies the coupling of the first oscillator (``controller'') with the second one (``target''). Let us introduce a dissipative process of the form $\Lrel{=}\sum_k\Lbd_{\nrcA_k}$ where $\nrcA_k{\defeq}e^{{-}i\kappa_k\hat x_1}\nrcf_k(\hat x_2)$ and $\nrcf_k(x_2)$ are real-valued functions. On the one hand, this process does not affect the evolution of the first moments of the target:
\begin{gather}\label{_nrc_master_eq2}
\tder{}t\midop{\hat x_2}{=}\frac{\midop{\hat p_2}}{m_2};~\tder{}{t}\midop{\hat p_2}{=}
{-}\midop{\tpder{U_2(\hat x_2)}{\hat x_2}}
{-}\midop{\hat x_1\tpder{\chi(\hat x_2)}{\hat x_2}}.
\end{gather}
On the other hand, $\Lrel$ exerts a translationally invariant statistical force on the controller. Indeed, the average value of any controller operator $\hat h{=}h(\hat p_1,\hat x_1)$ in the case $\kappa_k{\ll}\hbar^{-1}\sqrt{\midop{\hat H_1}m_1}$ evolves as
\begin{gather}\label{_nrc_master_eq1}
\tder{}t\midop{\hat h}{\simeq}\frac{i}{\hbar}\midop{[\hat H_1,\hat h]}{+}\midop{\ehat\nrcF\tpder{\hat h}{\hat p_1}}{+}\midop{\ehat\nrcD\tpder{^2\hat h}{\hat p_1^2}},
\end{gather}
where $\ehat\nrcF{=}{-}\chi(\hat x_2){-}\hbar\sum_k\kappa_k \nrcf_k^2(\hat x_2)$, and $\ehat\nrcD{=}\frac{\hbar^2}{2}{\sum_k\kappa_k^2\nrcf_k^2(\hat x_2)}$. Note that Eq.~\eqref{_nrc_master_eq1} is exact for $\hat h{=}\hat x,\hat p$. Consider the choice of parameters $\kappa_k$ and functions $\nrcf_k(x)$ such that $\ehat\nrcF{=}0$ and $\ehat\nrcD{=}$const [e.g{.},\ $\kappa_1{=}{-}\kappa_2$, $g_{k}(x){=}\sqrt{\frac1{2\hbar}(\max(|\kappa_k^{-1}\chi(x)|){-}\kappa_k^{-1}\chi(x))}$ ]. In this case, Eq.~\eqref{_nrc_master_eq1} no longer depends on  $\hat x_2$, and hence the controller acts  nonreciprocally on the target, as intended. 

The simplified variant of the proposed scheme can be implemented in the laboratory by merging the principles behind the sympathetic cooling and the plasmonic field enhancement in the setup shown in Fig.~\ref{@FIG.02b} and further specified in \appref{@APP:sympathetic_nrc}.
Following the same ideas, one can also nonreciprocally couple the electronic and translational degrees of freedom (see \appref{@APP:nonreciprocal_vibronic_coupling}).
 The detailed analysis will be presented in the forthcoming paper.

\section{Summary and outlook}
The history of quantum control \cite{2009-Brif} clearly illustrates how synergy of 
new technology and state-of-art theory gives birth to new rapidly growing research areas.
Today we still lack the theoretical and technological yeast 
to raise the scope of quantum control applications in chemistry by replacing current laser technologies with quantum reservoir engineering. At the same time, friction forces abound in nature have never been systematically considered for reservoir engineering. We have shown that these forces are powerful quantum control instruments despite the seemingly stringent no-go limitations. Moreover, the versatile experimental prototyping of the quantum friction effects is already feasible within the standard  methods of quantum optics. The example of nonreciprocal couplings has illustrated that quantum friction reservoir engineering can be advanced by borrowing ideas from quantum optics technology. In turn, Doppler cooling in quantum optics and cold atom physics can gain from optimizing the spectral properties of laser fields according to the presented quantum friction theory of optimal thermalization. We hope that this theory and the no-go theorems, which have remained as mere conjectures for over 40 years, may become a centerpiece of the long-standing puzzle to consistently introduce dissipative forces into quantum mechanics\footnote{Current status and persistent issues with the quantization of friction are reviewed in Refs.~\cite{2016-Bondar,1997-Kohen} (note errata \cite{2016-Bondar-a}). Challenges and controversies of the existing situation further highlighted in the discussions \cite{1998-Wiseman,2001-O'Connell,2001-Vacchini} of original works \cite{1998-Gao,2000-Vacchini}.}.
\begin{acknowledgements}
The authors thank the National Science Foundation (Award number CHEM-1012207 to T.S.) for support.
D.I.B. is supported by the 2016 AFOSR Young Investigator Research Program.
\end{acknowledgements}

\onecolumngrid
\bibsection
\begin{center}
\Large\protect{\texttt{\uppercase{Supplemental material}}}
\end{center}
\twocolumngrid
\appendix
\setcounter{section}{0}
\title{Quantum friction: 
environment engineering perspectives
\texorpdfstring{\\}\protect{\texttt{Supplemental material}}}

\author{Dmitry V. Zhdanov}
\email{dm.zhdanov@gmail.com}
\affiliation{Department of Chemistry, Northwestern University, 2145 Sheridan Road, Evanston, Illinois 60208-33113 USA}
\author{Denys I. Bondar}
\affiliation{Princeton University, Princeton, NJ 08544, USA}
\author{Tamar Seideman}
\email{t-seideman@northwestern.edu}
\affiliation{Department of Chemistry, Northwestern University, 2145 Sheridan Road, Evanston, Illinois 60208-33113 USA}
\maketitle
\section*{Introduction}
This supplemental material is organized as follows. In the Sections~\ref{@APP:theo:trans_inv_Lbd}, \ref{@APP:theo:no_go(T=0)} and \ref{@APP:theo:no_go(T>0)} we give the proofs of Lemma~\ref{@theo:trans_inv_Lbd}, first, and second no-go theorems, respectively. The supporting mathematical derivations for the Doppler cooling model discussed in the main text are provided in Section~\ref{@APP:phys_meaning}. Finally, Sections~\ref{@APP:sympathetic_nrc} and \ref{@APP:nonreciprocal_vibronic_coupling} detail the scheme of ``sympathetic'' nonreciprocal control over the translational motion of two atoms, and the method of nonreciprocal coupling between electronic (or spin) and vibrational degrees of freedom on the example of the two-level system (TLS) coupled with the harmonic oscillator.

The Roman numbers in parentheses refer everywhere to the equations in the main text of the letter.

\section{The proof of lemma~\ref{@theo:trans_inv_Lbd}\label{@APP:theo:trans_inv_Lbd}}
First, note that the property \eqref{Quantum_Liouville_trans_inv_nD} of the translational invariance can be equivalently reformulated as
\begin{gather}\label{Quantum_Liouville_trans_inv_nD(reformulated)}
\forall \dxx: {\cal R}_{\dxx}\Lrel{\cal R}_{-\dxx}{=}\Lrel,
\end{gather}
where
$
{\cal R}_{\dxx}{=}e^{{-}\frac{i}{\hbar}\dxx\hat\pp}\odot e^{\frac{i}{\hbar}\dxx\hat\pp}
$
is the superoperator of translational shift: $\forall g(\hat\xx): {\cal R}^{\intercal}_{\dxx}[g(\hat\xx)]{=}g(\hat\xx{+}\dxx)$.

With the help of the canonical commutation relations, any operator $\hat L_k{=}L_k(\hat\pp,\hat\xx)$ can expanded in the series
$\hat L_k{=}\sum_{l,m}c_{k,l,m}\hat B_{l,m}$, where $\hat B_{l,m}{=}e^{{-}i\set{\kappa}_l\hat\xx}g_m(\hat\pp)$ and the functions $g_m(\pp)$ constitute a set of (not necessarily orthogonal) basis functions. Using this expansion, any superoperator of form $\Lrel{=}\sum_k\Lbd_{\hat L_k}$ can be rewritten as
\begin{gather}
\Lrel{=}\sum_{k,l_1,m_1,l_2,m_2}c_{k,l_1,m_1}c_{k,l_2,m_2}^*\tilde{\cal L}^{\idx{lbd}}_{\hat B_{l_1,m_1},\hat B_{l_2,m_2}},
\end{gather}
where
\begin{gather}
\tilde{\cal L}^{\idx{lbd}}_{\hat A_1,\hat A_2}{\defeq}\hat A_1\odot\hat A_2^{\dagger}-\frac12(\hat A_1\hat A_2^{\dagger}\odot{+}\odot\hat A_1\hat A_2^{\dagger}).
\end{gather}
It follows from Eq.~\eqref{Quantum_Liouville_trans_inv_nD(reformulated)} that if $\Lrel$ is translationally invariant then it should satisfy the identity
\begin{gather}
\Lrel{=}\left.\frac{1}{(2L)^N}\int_{-L}^{L}...\int_{-L}^{L}\,{\cal R}_{\dxx}\Lrel{\cal R}_{-\dxx}\mathrm{d}^N\diff\dxx\right|_{L{\to}\infty}{=}\notag\\
\sum_{l}\sum_{m_1,m_2}\tilde c^{(l)}_{m_1,m_2}\tilde{\cal L}^{\idx{lbd}}_{\hat B_{l,m_1},\hat B_{l,m_2}},\label{Quantum_Liouville_trans_inv_nD_a}
\end{gather}
where the Hermitian matrices $\tilde c^{(l)}$ are defined as
\begin{gather}
\tilde c^{(l)}_{m_1,m_2}{=}\sum_{k}c_{k,l,m_1}c_{k,l,m_2}^*.
\end{gather}
Let us substitute in Eq.~\eqref{Quantum_Liouville_trans_inv_nD_a} the matrices $\tilde c^{(l)}$ with their Jordan decomposition $\tilde c^{(l)}{=}\tilde u^{(l)}\tilde\gamma^{(l)}{\tilde u^{(l)}}^{\dagger}$, where $\tilde u^{(l)}$ is unitary and $\tilde\gamma^{(l)}$ is diagonal. The result is
\begin{gather}\label{Quantum_Liouville_trans_inv_nD_b}
\Lrel{=}\sum_{l,m}{\cal L}^{\idx{lbd}}_{\hat A_{l,m}},
\end{gather}
where $A_{l,m}{=}\tilde f_{l,m}(\hat\pp)e^{{-}i\set{\kappa}_l\hat\xx}$ and $\tilde f_{l,m}(\hat\pp){=}\sqrt{\tilde\gamma^{(l)}_{m,m}}\times\sum_{m'}\tilde u^{(l)}_{m',m}g_{m'}(\hat\pp)$. Finally, note that Eq.~\eqref{Quantum_Liouville_trans_inv_nD_b} can be cast into the form \eqref{theorem:trans_inv_Lbd} by replacing the compound index $\{l,m\}$ with the single consecutive index $k$. The lemma is proven.
\begin{remark}
In this work, the Gaussian (continuous) translationally invariant dissipators of form 
\begin{gather}\label{_Lbd_Gaussian}
{\LvnG}{=}\sum_k\Lbd_{\ehat\AG_k},~~\ehat\AG_k{=}\kkappaG_k\hat\xx{+}\ffG_k(\hat\pp)~~(\kkappaG_k{\in}\mathbb{R}^N)
\end{gather}
are treated as the limiting case of Eq.~\eqref{theorem:trans_inv_Lbd} with $\kkappa_k{=}\epsilon\kkappaG_k{\to}0$. Specifically, one can verify by direct calculation that
\begin{gather}\label{_Lbd_Gaussian[lemma-form]}
\Lbd_{\ehat\AG_k}{=}\left.\Lbd_{\ehat\AG_{k,+}}{+}\Lbd_{\ehat\AG_{k,-}}\right|_{\epsilon{\to}0},~\ehat\AG_{k,\pm}{=}\tfrac1{\sqrt 2}\left(\tfrac{i}{\epsilon}{\pm}\ffG_k(\hat\pp)\right)e^{{\mp}i\epsilon\kkappaG_k\hat\xx}.
\end{gather}
\end{remark}
\section{The proof of no-go theorem~\ref{@theo:no_go(T=0)} (by contradiction)\label{@APP:theo:no_go(T=0)}}
Suppose that some eigenstate $\ket{\Psi_{0}}$ of Hamiltonian $\hat H$ simultaneously corresponds to the fixed point of the quantum Liouvillian $\cal L$ defined by Eqs.~\eqref{Quantum_Liouville_equation_nD} and \eqref{theorem:trans_inv_Lbd}. Denote as $\ket{\Psi_{\dxx}}$ the translationally displaced copy of $\ket{\Psi_{0}}$:
$\ket{\Psi_{\dxx}}{=}e^{{-}\frac{i}{\hbar}\dxx\hat\pp}\ket{\Psi_{0}}$.
%
%
Using the definition of the associated wavefunction $\Psi_{\dxx}(\pp)$ in the momentum space, one can write: $\ket{\Psi_{\dxx}}{=}\Psi_{0}(\hat\pp)\ket{\dxx}$, where $\ket{\dxx}$ is the eigenstate of position operator: $\hat x_k\ket{\dxx}{=}\delta x_k\ket{\dxx}$, $\scpr{\dxx}{\dxx'}{=}\delta(\dxx{-}\dxx')$. 

The linearity of $\Lrel$ together with the property \eqref{Quantum_Liouville_trans_inv_nD(reformulated)} of translational invariance imply that $\forall g(\xx'): \Lrel[\int g(\xx')\proj[\Psi_{\xx'}]{\Psi_{\xx'}}d^N \xx']{=}0$, which can be equivalently restated as
\begin{gather}\label{_w_g}
\forall g(\xx'): \Lrel[\hat w_g]{=}0,~~~\hat w_g{=}\Psi_{0}(\hat\pp)g(\hat\xx)\Psi_{0}(\hat\pp)^{\dagger}.
\end{gather}
Consider the case $g(\xx){=}g_{\llambda}(\xx){=}e^{-i\llambda\xx}$, where $\llambda$ is some real $N$-dimensional vector. Then, the condition \eqref{_w_g} reads:
\begin{gather}
\Lrel[\hat w_{g_{\llambda}}]{=}G(\hat\pp,\hat\pp{+}\hbar\llambda)\Psi_0 (\hat\pp)\Psi_0(\hat\pp{+}\hbar\llambda)^{\dagger}e^{-i \llambda\hat{\xx}}{=}0\label{_Lrel(w_g)},
\end{gather}
where
\begin{gather}
G(\pp,\pp'){=}\sum_k\left(
F_k(\pp)F_k(\pp')^{*}{-}\tfrac{|\tilde f_k(\pp)|^2{+}|\tilde f_k(\pp')|^2}2
\right),
\\
F_k(\pp){=}\tilde f_k (\pp{+}\hbar\kkappa_k) \frac{\Psi_0(\pp{+}\hbar\kkappa_k)}{\Psi_0(\pp) }.
\end{gather}
Condition \eqref{_Lrel(w_g)} implies that $\forall \pp,\pp'{\in}\mathbb{R}^N: G(\pp,\pp'){=}0$\footnote{Possibly, except a zero measure subset of points $\{\pp,\pp'\}$ where $\Psi_0(\hat\pp)\Psi_0(\hat\pp')^{\dagger}{=}0$.}. In particular, this means that 
\begin{gather}\label{dG/dp_1dp_2}
\begin{split}
\forall n,\forall \pp,\pp'{\in}\mathbb{R}^N:& \pder{^2}{p_{n}\partial p'_{n}} G(\pp,\pp'){=}\\
&\sum_k
\tpder{F_k^{(n)}(\pp)}{p_n}\left(\tpder{F_k^{(n)}(\pp')}{p_n'}\right)^{*}{=}0.
\end{split}
\end{gather} 
Equality \eqref{dG/dp_1dp_2} can be satisfied only if $\forall k: F_k(\pp){\propto}$const, i.e., if $\tilde f_k (\pp){=}c_k\frac{\Psi_0(\pp{-}\hbar\kkappa_k)}{\Psi_0(\pp)}$, where $c_k$ is some real constant. Substitution of this expression and $\llambda{=}0$ into Eq.~\eqref{_Lrel(w_g)} gives the following necessary condition for the asymptotic relaxation to the ground state:
\begin{gather}\label{condition_in_Psi_0}
\forall \pp{\in}\mathbb{R}^N: \sum_kc_k^2\left({|\Psi_0(\pp{-}\hbar\kkappa_k)}|^2{-}|{\Psi_0(\pp)}|^2\right)=0.
\end{gather}
\newcommand{\cchi}{\boldsymbol{\chi}}
Condition \eqref{condition_in_Psi_0} is equivalent to
\begin{gather}\label{condition_in_Psi_0_FT}
\forall \cchi{\in}\mathbb{R}^{N}: \phi(\cchi)\sum_kc_k^2\left(e^{-i\hbar\cchi\kkappa_k}{-}1\right){=}0,
\end{gather}
where $\phi(\cchi)$ denotes the Fourier transform of $|\Psi_0(\pp)|^2$. Equality \eqref{condition_in_Psi_0_FT}
can be satisfied for all $\cchi$ iif $\phi(\cchi)$ is nonzero only at the points where $\forall k: \hbar\cchi\kkappa_k\bmod{2\pi}{=}0$, i.e., only when $|\Psi_0(\pp)|^2$, and hence $U(\xx)$, are (quasi)periodic%
\footnote
{In the case of Gaussian dissipator \eqref{_Lbd_Gaussian} if follows from \eqref{dG/dp_1dp_2} that 
$\ffG(\pp){=}-i\hbar\kkappaG\pder{}{\pp}\ln(\Psi_0(\pp))$, and the 
condition \eqref{condition_in_Psi_0_FT} reduces to
\begin{gather}\label{condition_in_Psi_0_FT[Gaussian]}
\forall \cchi{\in}\mathbb{R}^{N}: {-}\phi(\cchi)\sum_k\left(\hbar\cchi\kkappaG_k\right)^2{=}0.\tag{\ref*{condition_in_Psi_0_FT}*}
\end{gather}
Similarly to \eqref{condition_in_Psi_0_FT}, the real part of the lhs of Eq.~\eqref{condition_in_Psi_0_FT[Gaussian]} is nonpositive, and the equality can be satisfied only if $\forall k: \kkappaG_k{=}\zzero$, i.e.\ only if $\forall k: \AG_k{=}0$. In other words, in the case of Gaussian dissipators \eqref{_Lbd_Gaussian} the statement of no-go theorem is valid for all potentials $U(\xx)$, including quasiperiodic ones.
}%
. 
This result completes the proof.


\section{The proof of no-go theorem~\ref{@theo:no_go(T>0)} (by contradiction)\label{@APP:theo:no_go(T>0)}}

\newcommand{\dx}{\delta x}
\newcommand{\rnorm}{\tilde N}

Denote as $\Psi_{k,\dxx}(\pp)$ and $E_k$ $(k{=}0,...,\infty)$ the momentum-space wavefunction and energy of the $k$-th eigenstate $\ket{\Psi_{k,\dxx}}$ of the displaced Hamiltonian $H(\hat\pp,\hat\xx{-}{\dxx})$. Then, $\ket{\Psi_{k,\dxx}}{=}\Psi_{k,\zzero}(\hat\pp)\ket{\dxx}$, where $\ket{\dxx}$ is the eigenstate of position operators: $\hat\xx\ket{\dxx}{=}\dxx\ket{\dxx}$, $\scpr{\dxx}{\dxx'}{=}\prod_{n{=}1}^{N}\delta(\dx_n{-}\dx_n')$.
The thermal state of the displaced system can be written in these notations as

\begin{gather}
\hat\rho_{\theta ,\dxx}=\rnorm\sum_ke^{{-}\frac{E_k}{\theta}}\Psi_{k,\zzero}(\hat\pp)\proj[\dxx]{\dxx}\Psi_{k,\zzero}(\hat\pp)^{\dagger},
\end{gather}
where $\rnorm{=}(\sum_ke^{{-}\frac{E_k}{\theta}})^{-1}$ is the normalization constant.
Suppose that there exists such relaxation superoperator of form \eqref{theorem:trans_inv_Lbd} that $\Lrel[\hat\rho_{\theta ,\zzero}]{=}0$. Then, the translational invariance of $\Lrel$ implies that
$\forall\dxx: \Lrel[\hat\rho_{\theta ,\dxx}]{=}0$. The later equality can be equivalently rewritten as
\begin{gather}\label{_Lrel(_w_{theta,g})=0}
\forall g(x): \Lrel[\hat w_{\theta ,g}]{=}0,
\end{gather}
where
\begin{gather}\label{_w_{theta,g}}
\hat w_{\theta,g}{=}\rnorm\sum_ke^{{-}\frac{E_k}{\theta}}\Psi_{k,\zzero}(\hat\pp)g(\hat\xx)\Psi_{k,\zzero}(\hat\pp)^{\dagger}.
\end{gather}

Consider the case $g(\xx){=}g_{\llambda}(\xx){=}e^{{-}i\llambda\xx}$, where $\llambda$ is some real $N$-dimensional vector. The result of application of $\Lrel$ to $\hat w_{\theta,g_{\llambda}}$ can be represented after some algebra as
\begin{gather}\label{__Lrel[w]}
\Lrel[\hat w_{\theta,g_{\llambda}}]{=}G_1\left(\hat\pp{+}\tfrac{\hbar\llambda}{2},\llambda\right)e^{{-}i\llambda\hat\xx},
\end{gather}
where
\begin{align}\label{__G_1(p,lambda)}
G_1(\pp,\llambda)&=\sum_k \biggl(Q_{k,n}(\pp{+}\hbar\kkappa_k,\llambda)-\frac{1}{2}\BFn_{\theta}(\pp,\llambda){\times}\notag\\ &\left(\left|\tilde{f}_k\left(\pp{+}\tfrac{\hbar \llambda }{2}\right)\right|^2{+}\left|\tilde{f}_k\left(\pp{-}\tfrac{\hbar \llambda}{2}\right)\right|^2\right)\biggr),
\\
Q_k(\pp,\llambda )&=\BFn_{\theta}(\pp,\llambda)\tilde{f}_k(\pp{-}\tfrac{\hbar \llambda }{2}) \tilde{f}_k^*(\pp{+}\tfrac{\hbar \llambda }{2}).
\end{align}
In derivation of \eqref{__G_1(p,lambda)} the identity
\begin{gather}
\BFn(\pp,\llambda ){=}
\rnorm\sum_k e^{{-}\frac{E_k}{\theta}}\Psi_{k,\zzero}(\pp{-}\tfrac{\hbar\llambda}{2})\Psi_{k,\zzero}^{*}(\pp{+}\tfrac{\hbar\llambda}{2})
\end{gather}
was used which follows directly from the definition \eqref{_Blokhintsev_function} of the Blokhintsev function.

Eqs.~\eqref{_Lrel(_w_{theta,g})=0} and \eqref{__Lrel[w]} require that 
\begin{gather}\label{__G_1(p,lambda){=}0}
\forall\pp,\llambda:G_1(\pp,\llambda){=}0,
\end{gather} 
and hence $\forall \llambda: \bar G_2(\llambda){=}\int_{{-}\infty}^{\infty}\ldots\int_{{-}\infty}^{\infty}\diff^N \pp\,G_2(\pp,\llambda){=}0$,
where
\begin{gather}\label{__G_2(p,lambda)}
\begin{split}
G_2(\pp,&\llambda){=}G_1(\pp,\llambda){+}G_1(\pp,{-}\llambda){=}
\\
\sum_k&\biggl\{
\sum_{\alpha,\beta{=}\pm1}\beta Q_k\left(\pp{+}\tfrac{\beta{+}1}2\hbar\kkappa_k,\alpha\llambda\right){-}
\\
&\left|\tilde{f}_k\left(\pp{+}\tfrac{\hbar\llambda }{2}\right){-}\tilde{f}_k\left(\pp{-}\tfrac{\hbar \llambda}{2}\right)\right|^2\BFn_{\theta}(\pp,\llambda)
\biggr\}.
\end{split}
\end{gather}
The last equality in \eqref{__G_2(p,lambda)} is obtained assuming that $\BFn_{\theta}(\pp,{-}\llambda){=}\BFn_{\theta}(\pp,\llambda)$ (see Eq.~\eqref{_B(p,lambda)-features-a'}).
It is easy to check that the integration over the first term in curly brackets in \eqref{__G_2(p,lambda)} cancels out, so that
\begin{align}
\bar G_2(\llambda){=}&{-}\int_{{-}\infty}^{\infty}\ldots\int_{{-}\infty}^{\infty}\diff^N \pp{\times}\notag\\
&\sum_k\left|\tilde{f}_k\left(\pp{+}\tfrac{\hbar\llambda }{2}\right){-}\tilde{f}_k\left(\pp{-}\tfrac{\hbar\llambda}{2}\right)\right|^2\BFn(\pp,\llambda).\label{_bar_G_2(p,lambda)}
\end{align}
According to the supposition \eqref{_B(p,lambda)-features-a'}, the integrand in \eqref{_bar_G_2(p,lambda)} is nonnegative. Moreover, $\bar G_2(\llambda){=}0$ iif $\forall k: \tilde f_k(\pp){=}c_k{=}$const. Hence, the expression \eqref{__G_1(p,lambda)} for $G_1(\pp,\llambda)$ can be simplified as
\begin{gather}\label{__G_1_simplified(p,lambda)}
G_1(\pp,\llambda )=\sum _k c_k^2 \left(\BFn\left(\pp{+}\hbar\kkappa_k,\llambda \right){-}\BFn(\pp,\llambda )\right).
\end{gather}
Note that the terms $\Lbd_{\hat A_k}$ in Eq.~\eqref{theorem:trans_inv_Lbd} with $\tilde f_k(\pp){=}$const will have non-trivial effect only if $\kkappa_k{\ne}0$%
\footnote
{In the case of Gaussian dissipator \eqref{_Lbd_Gaussian} Eq.~\eqref{__G_1_simplified(p,lambda)} reduces to
\begin{gather}\label{__G_1_simplified(p,lambda)[Gaussian]}
G_1(\pp,\llambda)=\frac12\hbar^2\sum_{k,m,n}\mu_{k,n}\mu_{k,m}\pder{^2}{p_np_m}\BFn(\pp,\llambda ).\tag{\ref*{__G_1_simplified(p,lambda)}*}
\end{gather}
By assumption \eqref{_B(p,lambda)-features-b'}, the quadratic form $\pder{^2}{p_np_m}\BFn(\pp,\llambda )$ in \eqref{__G_1_simplified(p,lambda)[Gaussian]} is negative-definite at $\{\pp,\llambda \}{=}\{\zzero,\zzero\}$. Hence, $G_1(\zzero,\zzero){<}0$, which contradicts Eq.~\eqref{__G_1(p,lambda){=}0} and completes the proof for this case.
}%
. However, it follows from \eqref{_B(p,lambda)-features-b'} that in this case $G_1(\zzero,\zzero){<}0$ which contradicts Eq.~\eqref{__G_1(p,lambda){=}0}. The theorem is proven.

\section{Doppler cooling as an example of quantum friction\label{@APP:phys_meaning}}

In this section, we provide the detailed analysis of the Doppler cooling example introduced in the main text of the letter (see Fig.~\ref{@FIG.01'} in the main text) and prove that the cooling mechanism is the quantum friction of form \eqref{theorem:trans_inv_Lbd-isotropic}. 

For the spatial arrangement depicted in Fig.~\ref{@FIG.01'} the translation motion of the atom along $x$-axis is coupled to the field-induced electron dynamics since each absorbed or coherently emitted photon changes the $x$-component of atomic momentum hereafter denoted as $p$. The master equation which describes this coupled dynamics can be written within the rotating wave approximation in the form \eqref{Quantum_Liouville_equation_nD} with  

\begin{gather}
\begin{split}
\hat H{=}&\frac{\hat p^2}{2m}{-}\hbar\omega_{\idx{a}}\proj{\es{g}}{+}
\biggl\{\xi_1(t)\proj[\es{e}_1]{\es{g}}e^{-i(\omega t{-}\kappa \hat x)}{+}\\&
\xi_2(t)\proj[\es{e}_2]{\es{g}}e^{-i(\omega t{+}\kappa \hat x)}{+}\mbox{h.c.}
\biggr\}
\end{split}
\end{gather}
and
\begin{gather}\label{doppler_L_rel}
\Lrel{=}{\gamma}\sum_{n{=}1}^2\Lbd_{\proj[\es{g}]{\es{e}_n}}.
\end{gather}
Here $\xi_k(t){=}{-}\frac12\vec d_k\vec{\cal E}_k(t)$, where $\vec d_1$ and $\vec d_2$ are the transition dipole moments associated with the $s{\to}p_z$ and $s{\to}p_y$ electronic transitions into degenerate electronically excited sublevels $\es{e}_1$ and $\es{e}_2$, respectively, and $\vec{\cal E}_k(t)$ is the slowly varying complex amplitude of the associated field component. The remaining notations are defined in the body of the letter.

\newcommand{\evsop}[2]{\mathop{{{\cal U}^{#1}_{#2}}}}
\newcommand{\evsopt}[2]{\mathop{{{\cal U}^{#1}_{#2}}^{\intercal}}}
\newcommand{\TOdir}{\stackrel{\Leftarrow}{{\cal T}}}
\newcommand{\TOinv}{\stackrel{\Rightarrow}{{\cal T}}}
\newcommand{\Pg}{\hat P_{\es{g}}}
The mean value of any observable of form $\hat O{=}f(\hat p,\hat x)$ can be written in Heisenberg representation as:
\begin{gather}\label{_<O>}
\midop{\hat O(t)}{=}\Tr[\hat\rho_0\evsopt{\cal L}{t,t_0}[\hat O]],
\end{gather}
where we define:
\begin{gather}\label{_evsop}
\forall {\cal L}(t): \evsop{\cal L}{t,t_0}\stackrel{\idx{def}}{=}\TOdir e^{\int_{t{=}t_0}^{t}{\cal L}\diff t}.
\end{gather} 
The symbol $\TOdir$ in \eqref{_evsop} denotes the chronological ordering superoperator which arranges operators in direct (inverse) time order for $t{>}t_0$ ($t{<}t_0$). Let us also define the following notations for the interaction representation generated by arbitrary splitting ${\cal L}(t){=}{\cal L}_0+{\cal L}_1(t)$:
\begin{gather}\label{superoperator_interaction_representation}
(\evsop{\cal L}{t,0})^{\intercal}{=}\evsop{(\cal L_0^{\intercal})}{t,0}\evsop{({\cal L}_{\idx{I}}^{\intercal})}{t,0},
\end{gather} 
where the interaction Liouvillian reads
\begin{gather}\label{interaction_Liouvillian}
{\cal L}_{\idx{I}}^{\intercal}(\tau){=}{\evsop{({\cal L}^{\intercal}_0)}{{-}\tau,0}}{\cal L}_1^{\intercal}(t{-}\tau){\evsop{({\cal L}_0^{\intercal})}{\tau,0}}.
\end{gather}
In the case ${\cal L}_0'{=}\frac{-i}{\hbar}[\frac{\hat p^2}{2m}{-}\hbar\omega_{\idx{a}}\proj{\es{g}},\odot]$ the associated interaction liouvillian \eqref{interaction_Liouvillian} in the rotating wave approximation takes the form:
\begin{gather}\label{L_I'-}
{\cal L}_{\idx{I}}'{\simeq}\frac{-i}{\hbar}[\hat H',\odot]{+}\sum_{n{=}1}^2\Lbd_{\proj[\es{g}]{\es{e}_n}},
\end{gather}
where
\begin{gather}
\hat H'(\tau){=}\sum_{n{=}1}^2\hat\chi_n(\tau)\proj[\es{g}]{\es{e}_n}{+}\mbox{h.c.};\\
\hat\chi_1(\tau){=}\xi_1^*(t{-}\tau)e^{i(\omega t{-}\kappa \hat x{-}(\Delta{-}\frac{\kappa \hat{p}}{m})\tau) };\\
\hat\chi_2(\tau){=}\xi_2^*(t{-}\tau)e^{i(\omega t{+}\kappa \hat x{-}(\Delta+\frac{\kappa \hat{p}}{m})\tau)},
\end{gather}
and $\Delta{=}\omega{-}\omega_{\idx{a}}$ is detuning of carrier frequency of radiation from atomic resonance in the case of system at rest.
Repeated application of the transformation \eqref{superoperator_interaction_representation} to \eqref{L_I'-} with ${\cal L}_0''{=}\Lrel{=}\sum_{n{=}1}^2\Lbd_{\proj[\es{g}]{\es{e}_n}}$ leads to expression:
\begin{gather}
(\evsop{\cal L}{t,0})^{\intercal}{=}\evsop{{\cal L_0'}^{\intercal}{+}\Lrel^{\intercal}}{t,0}\evsop{({{\cal L}_{\idx{I}}''}^{\intercal})}{t,0},
\end{gather}
so that 
\begin{gather}
\midop{\hat O(t)}{=}\Tr[(\evsop{{\cal L_0'}{+}\Lrel}{t,0}[\hat\rho_0])\evsop{({{\cal L}_{\idx{I}}''}^{\intercal})}{t,0}[\hat O]]\stackrel{t{\gg}\gamma^{-1}}{=}\\
\Tr[\Pg{(\evsop{{\cal L_0'}{+}\Lrel}{t,0}[\hat\rho_0])}\Pg({\evsop{({{\cal L}_{\idx{I}}''}^{\intercal})}{t,0}[\hat O]})\Pg]\label{<O(t)>-doppler(t->inf)},
\end{gather}
where $\Pg{=}\proj{\es{g}}$ and the last equality is due to the exponential damping of excited states populations induced by  relaxation superoperator \eqref{doppler_L_rel}.
Let us consider the evolution $\hat O(t)$ generated by the superoperator $\evsop{{{\cal L}_{\idx{I}}''}^{\intercal}}{t+\delta t,t}$:
\begin{gather}\label{generator_2-order-expansion}
\begin{split}
\hat O(t{+}&\delta t){=}\biggl(1{+}\int_t^{t{+}\delta t}{{\cal L}_{\idx{I}}''}^{\intercal}(\tau)\diff\tau{+}\\
&\int_t^{t{+}\delta t}\diff\tau_2\int_t^{\tau_2}d\tau_1{{\cal L}_{\idx{I}}''}^{\intercal}(\tau_2){{\cal L}_{\idx{I}}''}^{\intercal}(\tau_1)\biggr)\hat O(t).
\end{split}
\end{gather}
Integrands in Eq.~\eqref{generator_2-order-expansion} include the terms oscillating at frequencies $|\Delta{\pm}\frac{k \midop{\hat{p}}}m|$. In sequel we will consider the so-called weak-field limit when these oscillations are rapid relative to the characteristic timescales of the relevant processes, so that the contributions of the associated terms asymptotically vanish. In this limit, the second term in rhs of Eq.~\eqref{generator_2-order-expansion} disappears. The remaining terms constitute two decoupled evolution equations for the reduced density matrices $f_{\es{x}}(\hat p,\hat x,t{+}\delta t){=}\matel{\es{x}}{\hat O(t)}{\es{x}}$ ($\es{x}{=}\es{g},\es{e}$):
\begin{gather}\label{f_g(t)-func}
\begin{split}
f_{\es{g}}(\hat p,&\hat x,t{+}\delta t){=}\biggl({\odot}{+}\frac{1}{\hbar^2}\int_t^{t{+}\delta t}\diff\tau_2\int_t^{\tau_2}d\tau_1e^{\frac{1}{2} \gamma  (\tau_1{-}\tau_2)}\times\\
&\sum_{n{=}1}^2\biggl\{\hat\chi_n(\tau_2){\odot}{\hat\chi_n^{\dagger}(\tau_1)}{+}{\hat\chi_n(\tau_1)}{\odot}\hat\chi_n^{\dagger}(\tau_2){-}\\
&{\odot}{\hat\chi_n(\tau_1)}{\hat\chi_n^{\dagger}(\tau_2)}{-}\hat\chi_n(\tau_2) {\hat\chi_n^{\dagger}(\tau_1)}{\odot}\biggr\}\biggr)[f_{\es{g}}(\hat p,\hat x,t)];
\end{split}\\
f_{\es{e}}(\hat p,\hat x,t{+}\delta t){=}{\cal G}[f_{\es{e}}(\hat p,\hat x,t)]
\end{gather}
The explicit form of $\cal G$ is irrelevant for the sequel in view of Eq.~\eqref{<O(t)>-doppler(t->inf)}. The first two terms in the curly brackets in Eq.~\eqref{f_g(t)-func} can be transformed as
\begin{subequations}\label{f_g(t)-term}
\begin{gather}
\begin{split}\label{f_g(t)-1-st_term}
\hat\chi_1&(\tau_2){f_{\es{g}}(\hat p,\hat x,t)}{\hat\chi_1^{\dagger}(\tau_1)}{=}\\
&\xi_1^*(t{-}\tau_2)\xi_1(t{-}\tau_1)f_{\es{g}}(\hat{p}{+}\hbar\kappa,\hat{x}{+}\tfrac{\hbar\kappa}{m}\tau_2,t)e^{i \Delta_1(\hat p)(\tau_1{-}\tau_2)}{=}\\
&\xi_1^*(t{-}\tau_2)\xi_1(t{-}\tau_1)e^{i\hat\Delta_1(\hat p)(\tau_1{-}\tau_2) }f_{\es{g}}(\hat{p}{+}\hbar\kappa,\hat{x}{+}\tfrac{\hbar\kappa}{m}\tau_1,t),
\end{split}
\end{gather}
\begin{gather}
\begin{split}\label{f_g(t)-2-nd_term}
\hat\chi_1&(\tau_1){f_{\es{g}}(\hat p,\hat x,t)}{\hat\chi_1^{\dagger}(\tau_2)}{=}\\
&\xi_1(t{-}\tau_2)\xi_1^*(t{-}\tau_1)f_{\es{g}}(\hat{p}{+}\hbar\kappa,\hat{x}{+}\tfrac{\hbar\kappa}{m}\tau_1,t)e^{{-}i \Delta_1(\hat p)(\tau_1{-}\tau_2)}{=}\\
&\xi_1(t{-}\tau_2)\xi_1^*(t{-}\tau_1)e^{{-}i\hat\Delta_1(\hat p)(\tau_1{-}\tau_2) }f_{\es{g}}(\hat{p}{+}\hbar\kappa,\hat{x}{+}\tfrac{\hbar\kappa}{m}\tau_2,t),
\end{split}
\end{gather}
\end{subequations}
where
$
\Delta_1(p){=}\Delta{-}\frac{\kappa( p{+}\frac{\hbar\kappa}{2})}{m}.
$
The extra displacements $\frac{\hbar\kappa}{m}\tau_n$ in the $x$-dependencies of $f_{\es{g}}$ in Eqs.~\eqref{f_g(t)-term} account for the change of the velocity of atom after the photon absorption.
These displacements are typically very small compared to the characteristic scales of spatial change of the function $f_{\es{g}}$ and can be neglected. With this approximation, the exponentials and functions $f_{\es{g}}$ in Eqs.~\eqref{f_g(t)-term} commute, which allows to write:
\begin{subequations}\label{approximation_for_f_g(t)-term}
\begin{gather}
\begin{split}
\frac{1}{\hbar^2}\int_t^{t{+}\delta t}&\diff\tau_2\int_t^{\tau_2}d\tau_1e^{\frac{1}{2}\gamma(\tau_1{-}\tau_2)}{\times}\\
&\left(\hat\chi_1(\tau_2){\odot}{\hat\chi_1^{\dagger}(\tau_1)}{+}{\hat\chi_1(\tau_1)}{\odot}\hat\chi_1^{\dagger}(\tau_2)\right)[f_{\es{g}}(\hat{p},\hat{x},t)]{\simeq}\\
&2C_{+}(\hat p,t)f_{\es{g}}(\hat{p}{+}\hbar\kappa,\hat{x},t)C_{+}(\hat p,t)\delta t,
\end{split}\\
\begin{split}
\frac{1}{\hbar^2}\int_t^{t{+}\delta t}&\diff\tau_2\int_t^{\tau_2}d\tau_1e^{\frac{1}{2}\gamma(\tau_1{-}\tau_2)}{\times}\\
&\left(\hat\chi_2(\tau_2){\odot}{\hat\chi_2^{\dagger}(\tau_1)}{+}{\hat\chi_2(\tau_1)}{\odot}\hat\chi_2^{\dagger}(\tau_2)\right)[f_{\es{g}}(\hat{p},\hat{x},t)]{\simeq}\\
&2C_{-}(\hat p,t)f_{\es{g}}(\hat{p}{-}\hbar\kappa,\hat{x},t)C_{-}(\hat p,t)\delta t,
\end{split}
\end{gather}
\end{subequations}
where
\begin{subequations}\label{formulas_for_C+-}
\begin{widetext}
\begin{align}\label{-C+-}
C_{+}(p,t)&{=}
\sqrt{s_+(p){+}s_+^{*}(p)},   &s_+(p)&{=}\frac{1}{2\hbar^2\delta t}\int_t^{t{+}\delta t}\diff\tau_2\int_t^{\tau_2}d\tau_1\xi_1^*(t{-}\tau_2)\xi_1(t{-}\tau_1)e^{(i\Delta_1(p){+}\frac{\gamma}2)(\tau_1{-}\tau_2)},
\\
C_{-}(p,t)&{=}\sqrt{s_-(p){+}s_-^{*}(p)},   & s_{-}(p)&{=}\frac{1}{2\hbar^2\delta t}\int_t^{t{+}\delta t}\diff\tau_2\int_t^{\tau_2}d\tau_1\xi_2^*(t{-}\tau_2)\xi_2(t{-}\tau_1)e^{(i\Delta_1({-}p){+}\frac{\gamma}2)(\tau_1{-}\tau_2)}.
\end{align}
\end{widetext}
\end{subequations}
Substitution of approximations \eqref{approximation_for_f_g(t)-term} into \eqref{f_g(t)-func} gives:
\begin{gather}\label{effective_Liouvillian}
f_{\es{g}}(\hat p,\hat x,t{+}\delta t){=}\evsop{{\cal L}_{\idx{eff}}^{\intercal}}{t{+}\delta t,t}[f_{\es{g}}(\hat p,\hat x,t)],
\end{gather}
where
\begin{gather}
{\cal L}_{\idx{eff}}(t){=}{-}\frac{i}{\hbar}[\hat H_{\idx{eff}},\odot]{+}\Lrel^{\idx{eff}},
\end{gather}
\begin{gather}
\label{effective_friction}
\Lrel^{\idx{eff}}{=}
\Lbd_{C_{+}(\hat p,t)e^{i\kappa \hat x}}{}+\Lbd_{C_{-}(\hat p,t)e^{{-}i\kappa \hat x}},
\end{gather}
\begin{gather}
\hat H_{\idx{eff}}{=}i\hbar\sum_{m=\pm}(s_m(\hat p)-s_m^{*}(\hat p)).
\end{gather}
Eq.~\eqref{effective_Liouvillian} allows to calculate the averaging in \eqref{<O(t)>-doppler(t->inf)} within the reduced Hilbert space which involves only the translational degree of freedom:
\begin{gather}
\midop{\hat O(t)}\stackrel{t{\gg}\gamma^{-1}}{=}\Tr[\hat\rho_0^{\idx{red}}{\evsop{\frac{i}{\hbar}[\frac{\hat p^2}{2m},\odot]}{t,0}\evsop{{\cal L}_{\idx{eff}}^{\intercal}}{t,0}[\hat O]}]_{\idx{spatial}}\label{<O(t)>-spatial_only}.
\end{gather}
Here $\hat\rho_0^{\idx{red}}{=}\Tr[\hat\rho]_{\idx{el}}$ whereas $\Tr[\odot]_{\idx{el}}$ and $\Tr[\odot]_{\idx{spatial}}$ denote the partial traces over the electronic and translational subsystems.

The dissipator \eqref{effective_friction} reduces to the isotropic friction of form \eqref{theorem:trans_inv_Lbd-isotropic} provided that
\begin{gather}
\forall p:C_{+}({-}p,t){=}C_{-}(p,t){=}\fciso(p).
\end{gather}
It is easy to verify that this condition is realized in two important cases.
\subsection{Coherent laser driving}
In this regime, $\xi_1(t){=}\xi_2(t){=}\xi{=}$const, and there exists such $\delta t$ in \eqref{approximation_for_f_g(t)-term} that $\gamma^{-1}{
\gg}\delta t{\gg}\Delta_1^{-1}(p)$. Thence, the integrals in \eqref{formulas_for_C+-} can be easily computed, which gives:
\begin{gather}\label{_coherent_C_{+-}}
\Lrel^{\idx{eff}}{=}\Bdn_{\kappa,\fciso}^{\hat x,\hat p},~~
\fciso(p){=}|\xi|{\frac{\sqrt{\gamma}}{\hbar}\frac{\Delta_1(-p)}{(\frac{\gamma}2)^2{+} \Delta_1^2(-p)}},\\
\hat H_{\idx{eff}}{=}|\xi|^2\frac{\Delta_1(p)\left((\frac{\gamma}2)^2{-} \Delta_1^2(p)\right)}{\hbar\left((\frac{\gamma}2)^2+ \Delta_1^2(p)\right)^2}.
\end{gather}
Note what the Hamiltonian $\hat H_{\idx{eff}}$ describes the effect of the optical quadratic Stark shift which also can induce the effective potential forces on the system in the case of spatially non-uniform fields $\xi{=}\xi(x)$. 
\subsection{Incoherent driving}
Suppose that the the atom is illuminated by the two classical light sources with the equal spectral densities $I(\omega)$ at the atomic site and having coherence times in the range $\Delta_1^{-1}(p){\ll}t_{\idx{coh}}{\ll}\gamma^{-1}$. In this case, $\xi_{1}(t)$ and $\xi_{2}(t)$ represent the uncorrelated stationary stochastic processes. This allows one to choose such $\delta t$, that $\gamma^{-1}{\gg}\delta t{\gg}t_{\idx{coh}}$, and calculate the integrals in Eqs.~\eqref{formulas_for_C+-} neglecting the terms $\frac{\gamma}2$ in the exponents, which gives
\begin{gather}
\Lrel^{\idx{eff}}{=}\Bdn_{\kappa,\fciso}^{\hat x,\hat p},~~
\fciso(p){=}\frac{\pi d}{\hbar}\sqrt{\frac{1}{2 c}I(\omega{+}\Delta_1(-p))},
\end{gather}
where $I(\omega)$ is the spectral density of each beam. Also, here we assumed equal transition dipole momenta: $d{=}|\vec d_1|{=}|\vec d_2|$.

\section{``Sympathetic'' nonreciprocal control\label{@APP:sympathetic_nrc}}
Here we discuss the possible laboratory implementation of the simplified version of the nonreciprocal coupling scheme presented in the main text. In this scheme both the target and controller atoms as well as the metal nanoparticle are coaxially aligned along the $x$ axis and irradiated by the linearly polarized laser propagating antiparallel to the same axis, as shown in Fig.~\ref{@FIG.02b}. We assume the quadratic antibonding (repelling) atom-atom interaction of form
\begin{gather}
\hat H_{\idx{I}}{=}{-}\chi^{(0)}\hat x_1\hat x_2.
\end{gather} 
Let us choose the laser frequency $\omega$ to be off-resonant for the target atom but nearly resonant (with detuning $\Delta$) with $\es{g}\LR\es{e}$ electron transition in the controller atom. The field effect on the controller spatial motion can be calculated using the same procedure as in the case of Doppler cooling, Sec.~\ref{@APP:phys_meaning} which gives the following dissipative contribution to the quantum Liouvillian (cf. \eqref{_coherent_C_{+-}}):
\begin{gather}\label{_Lrel_sympathetic_gen}
\Lrel^{\idx{eff}}{=}{\Lbd_{\nrcf(\hat x_1,\hat x_2)e^{-i\kappa\hat x_1}}},~
\nrcf(x_1,x_2){=}|\xi|{\frac{\sqrt{\gamma}}{\hbar}\frac{\Delta}{(\frac{\gamma}2)^2{+} \Delta^2}}.
\end{gather}
Here $\kappa{=}\omega/c$, $\xi{=}{-}\frac12{\cal E(x_1)}d$, where $d$ is the value of the $\es{g}\LR\es{e}$ transition dipole moment, and $\gamma$ is the decay rate of the excited state $\es{e}$ (as in Sec.~\ref{@APP:phys_meaning}, for simplicity, we assume the case of non-radiative $\es{e}{\to}\es g$ decay). For typical transitions the photon momentum $\hbar\kappa$ is much smaller than atomic one. Assuming additionally that the value of $\gamma^{-1}$ is small compared to characteristic time of atomic motion, the effect of laser can be described in terms of uniform radiational pressure when the contribution of the last term $\midop{\ehat\nrcD\tpder{^2\hat h}{\hat p_1^2}}$ in \eqref{_nrc_master_eq1} is small compared to the first two terms, so one can set $\ehat\nrcD{\simeq}0$. For nonreciprocal coupling, such that $\ehat\nrcF{\simeq}0$, we additionally need $\nrcf(x_1,x_2)$ to be of special form
\begin{gather}\label{_Lrel_sympathetic_nrc}
\nrcf(x_1,x_2){=}\fc^{(0)}+\sqrt{\frac{\chi^{(0)}}{\hbar\kappa}}x_2+o(x_1,x_2).
\end{gather}
In \eqref{_Lrel_sympathetic_gen} we can neglect the weak dependence of $\fc$ on the atomic momentum $p_1$ but should account for 1) level shifts due to presence of the controller atom which result in position-dependent detuning $\Delta{=}\Delta(x_2{-}x_1)$, and 2) spatial dependence of laser field (and hence, the value of $\xi$) due to plasmon effect of nanoparticle: $\xi{=}\xi(x_1)$. Straightforward calculation shows that the relation \eqref{_Lrel_sympathetic_nrc} can be reduced to two conditions
\begin{subequations}
\begin{gather}\label{_symp_rec_1}
{\left.\pder{\log \xi(x_1)}{x_1}\right|_{x_1{=}0}}{=}\frac{\Delta(0) \left(\Delta(0)^2{-}\tfrac34 {\gamma}^2\right)}{(\tfrac{\gamma}{2})^4{-}\Delta(0)^4}\left.\pder{\Delta(r)}{r}\right|_{r{=}0},\\
\label{_symp_rec_2}
\xi (0){=}\sqrt{\frac{\chi^{(0)}}{\hbar\kappa}}\frac{\hbar  \left((\tfrac{\gamma}{2})^2+\Delta (0 )^2\right)^2}{\sqrt{\gamma } \left((\tfrac{\gamma}{2})^2{-}\Delta (0 )^2\right)\left.\pder{\Delta(r)}{r}\right|_{r=0}}.
\end{gather}
\end{subequations}
The first condition can be achieved in two ways: via tuning the lhs of Eq.~\eqref{_symp_rec_1} by changing the distance between the nanoparticle and controller atom or by varying $\Delta(0)$ in the rhs via adjusting the laser frequency $\omega$. Finally, the condition \eqref{_symp_rec_2} returns the magnitude of the required laser field. 

\section{Nonreciprocal vibronic coupling\label{@APP:nonreciprocal_vibronic_coupling}}
In this section we will consider the quantum system consisting of the coupled two-level system (TLS) and harmonic oscillator. Our aim is to nonreciprocally decouple the ``controller'' harmonic mode from the ``target'' TLS. 

The specific experimental arrangement which we are going to consider resembles the Doppler cooling experiment considered in the main text of letter except for now we will assume the constrained spatial motion in the potential well $U(x)$ and the case of $z$-polarized light propagating along axis $x$, so that only one electronic sublevel $\es{e}_1$ can be excited. Our Hamiltonian of interest (in interaction representation and after applying the rotating wave approximation) has the following form:

\begin{figure}[t!]
\centering\includegraphics[width=0.4\textwidth]{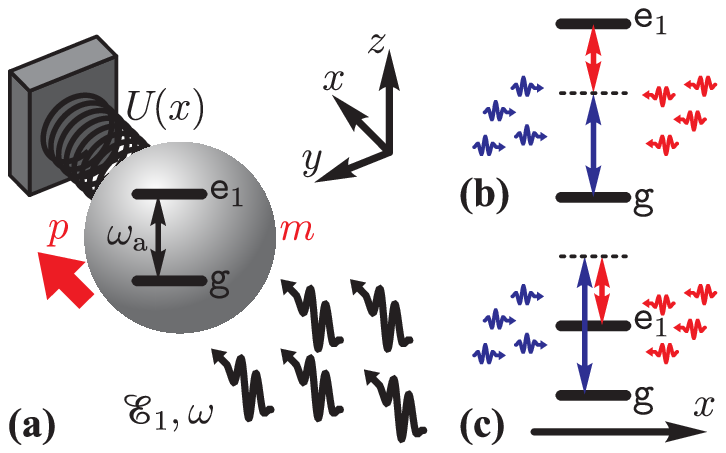}
\caption{(a) Physical implementation of the nonreciprocal vibronic coupling with dissipative term of form \eqref{nevc-Lrel}, \eqref{nevc-L+-[1]}. (b)(c) Examples of control over magnitude of $\kappa$ using nonlinear interactions: $\kappa_{\idx{eff}}{<}\kappa_0$ (b) and $\kappa_{\idx{eff}}{>}\kappa_0$ (c).\label{@FIG.03'}}
\end{figure}

\begin{gather}\label{nevc-Hamiltonian}
\hat H{=}\hat H_{\idx{el}}{+}\hat H_{\idx{vib}}{+}\hat H_{\idx{cpl}},
%
%
\end{gather}
where $\hat H_{\idx{vib}}{=}\frac{\hat p^2}{2m}{+}U(\hat x)$ describes the vibrational dynamics, $\hat H_{\idx{el}}$ is bare Hamiltonian of TLS and 
\begin{gather}
\hat H_{\idx{cpl}}{=}\chi\hat\sigma_3\hat x
\end{gather}
is the vibronic coupling term. Here the operators $\hat \sigma_k$ denote Pauli matrices in the basis $\{\es{g},\es{e}_1\}^{\intercal}$ of electronic states ($\hat \sigma_0$ states for identity matrix), and the rest of notations have the same meaning as in Section~\ref{@APP:phys_meaning}). Without loss of generality, we will further assume the case $\chi{>}0$. 
Consider the dissipation term of form 
\begin{gather}\label{nevc-Lrel}
\Lrel{=}\sum_{\alpha={\pm}}\Gamma\Lbd_{\hat L_{\alpha}},
\end{gather}
where $\hat L_{\pm}$ are defined by either of the following two formulas:
\begin{subequations}
\label{nevc-L+-}
\begin{gather}\label{nevc-L+-[1]}
\hat L_{\pm}{=}\frac12(\sigma_{1}{\mp}i\sigma_{2})e^{{\pm}i\kappa\hat x},\\
\hat L_{\pm}{=}\frac12(\sigma_{0}{\mp}\sigma_{3})\hat\sigma_{\pm}e^{{\pm}i\kappa\hat x}.\label{nevc-L+-[2]}
\end{gather}
\end{subequations}
Note that the dissipation of form \eqref{nevc-L+-[1]} can be realized in the Doppler cooling framework developed in Section~\ref{@APP:phys_meaning} with the following changes: a) only one broadband $z$-polarized incoherent radiation source is present; b) non-radiative decay can be neglected ($\gamma{=}0$). The corresponding possible experimental setup is shown in Fig.~\ref{@FIG.03'}a.
The Ehrenfest relations describing the dynamics of electronic and vibrational subsystems read:
\begin{subequations}\label{nevc-all}
\begin{gather}
\begin{split}
\der{}{t}&\midop{\hat f(\hat p,\hat x)}{=}\tfrac{i}{\hbar}[\hat H_{\idx{vib}}{+}(1{-}\Gamma\tfrac{\hbar\kappa}{\chi})\hat H_{\idx{cpl}},f(\hat p,\hat x)]{+}\\
&\Gamma\sum_{k{=}1}^{\infty}\biggl\{\frac{(\hbar\kappa)^{2n}}{(2n)!}\pder{^{2n}f(\hat p,\hat x)}{\hat p^{2n}}{+}\hat\sigma_3\frac{(\hbar\kappa)^{2n{+}1}}{(2n{+}1)!}\pder{^{2n{+}1}f(\hat p,\hat x)}{\hat p^{2n{+}1}}\biggr\},
\end{split}\label{nevc-d/dt(f(p,x))}\\
\der{}{t}\midop{\hat\sigma_k}{=}\frac{i}{\hbar}\midop{[\hat H_{\idx{el}}{+}\hat H_{\idx{cpl}},\hat\sigma_k]}{-}(1{+}\alpha\delta_{k,3}){\Gamma}\midop{\sigma_k},\label{nevc-d/dt(sigma)}
\end{gather}
\end{subequations}
where $\alpha{=}{+}1$ and $-1$ match the cases \eqref{nevc-L+-[1]} and \eqref{nevc-L+-[2]}, respectively. 
We can see that if one will set $\Gamma{=}\frac{\chi}{\hbar\kappa}$ and choose sufficiently small $\kappa$ then the dependence of Eq.~\eqref{nevc-d/dt(f(p,x))} on $\hat H_{\idx{cpl}}$ cancels out whereas the terms in curly brackets asymptotically vanish in the limit $\kappa{\to}0$ (note also that these terms are absent in the case of the first momenta $f(\hat p, \hat x){=}\hat p$ and $f(\hat p, \hat x){=}\hat x$). Hence, such choice corresponds to complete controller-target decoupling. 

Note, however, that in the limit $\kappa{\to}0$ the electron dynamics is dominated by the last term in Eq.~\eqref{nevc-d/dt(sigma)} since $\Gamma{=}\frac{\chi}{\hbar\kappa}{\to}\infty$. This implies the complete decoherence ($\hat \rho_{\idx{el}}{\to}\hat \sigma_0$) in the case \eqref{nevc-L+-[1]} and quantum Zeno effect (with measured operator $\hat\sigma_3$) for the choice \eqref{nevc-L+-[1]}. For this reason, the intermediate values of $\kappa$ are preferable which balance the effects of shot noise on both electronic and vibrational dynamics. However, the control over $\kappa$ is complicated by the fact that for the effective interaction the carrier frequency of radiation should be close to TLS transition frequency: $\omega_{\idx{}}{\simeq}|\Delta|$, which implies $\kappa{\simeq}\kappa_0{=}|\Delta|/c$. This restriction on $\kappa$ can be relaxed by employing the nonlinear interactions. For example, in order to increase the effective value of $\kappa$ one can use two incoherent photon sources aligned as shown in Fig.~\ref{@FIG.03'}c and having the carrier frequencies $\omega_{\idx{}}^{\idx{left}}$ and $\omega_{\idx{}}^{\idx{right}}$ satisfying the two-photon resonance condition $\omega_{\idx{a}}{\simeq}\omega_{\idx{}}^{\idx{left}}{-}\omega_{\idx{}}^{\idx{right}}$. In this case, $\kappa_{\idx{eff}}{\simeq}\frac{\omega_{\idx{}}^{\idx{left}}{+}\omega_{\idx{}}^{\idx{right}}}c{>}\kappa_0$. In similar fashion, one can use the two-photon transitions to achieve $\kappa_{\idx{eff}}{<}\kappa_0$, as shown in Fig.~\ref{@FIG.03'}b.


\begin{thebibliography}{99}
     \bibitem{BOOK-Razavy} M. Razavy, \emph{Classical and quantum dissipative systems} 
        (World Scientific, 2005).


     \bibitem{1976-Lindblad}{ G. Lindblad}, 
        ``On the Generators of Quantum Dynamical Semigroups,''
        Comm. Math. Phys. \textbf{48}, 119 (1976).


     \bibitem{1976-Lindblad-a}{ G. Lindblad}, 
        ``Brownian Motion of a Quantum Harmonic Oscillator,''
        Rep. Math. Phys. \textbf{10}, 393 (1976).

     \bibitem{1985-Dodonov}{ V. V. Dodonov and O. V. Manko}, 
        ``Quantum Damped Oscillator in a Magnetic Field,''
        Physica A \textbf{130}, 353 (1985).

     \bibitem{1997-Kohen}{ D. Kohen, C. C. Marston, and D. J. Tannor}, 
        ``Phase Space Approach to Theories of Quantum Dissipation,''
        J. Chem. Phys. \textbf{107}, 5236 (1997).

     \bibitem{2011-Krauter}{ H. Krauter, C. A. Muschik, K. Jensen, W. Wasilewski, J. M. Petersen, J. I. Cirac, and E. S. Polzik},
        ``Entanglement Generated by Dissipation and Steady State Entanglement of Two Macroscopic Objects,''
         Phys. Rev. Lett. \textbf{107}, 080503 (2011);
         { C. A. Muschik, H. Krauter, K. Jensen, J. M. Petersen, J. I. Cirac, and E. S. Polzik},
        ``Robust Entanglement Generation by Reservoir Engineering,''
         J. Phys. B: At. Mol. Opt. Phys. \textbf{45}, 124021 (2012).


     \bibitem{2012-Muller}{ Markus M\"uller, Sebastian Diehl, Guido Pupillo, and Peter Zoller},
        ``Engineered Open Systems and Quantum Simulations with Atoms and Ions,''
         in \textit{Advances In Atomic, Molecular, and Optical Physics} (Elsevier, 2012), p.~1.

     \bibitem{2014-Kronwald}{ A. Kronwald, F. Marquardt, and A. A. Clerk}, 
        ``Dissipative Optomechanical Squeezing of Light,''
        New J. Phys. \textbf{16}, 063058 (2014).

     \bibitem{2012-Eremeev}{ Vitalie Eremeev, Victor Montenegro, and Miguel Orszag},
        ``Thermally Generated Long-Lived Quantum Correlations for Two Atoms Trapped in Fiber-Coupled Cavities,''
         Phys. Rev. A \textbf{85}, 032315 (2012).
     \bibitem{2010-Pielawa}{ Susanne Pielawa, Luiz Davidovich, David Vitali, and Giovanna Morigi},
        ``Engineering Atomic Quantum Reservoirs for Photons,''
         Phys. Rev. A \textbf{81}, 043802 (2010).
     \bibitem{2012-Koga}{K. Koga and N. Yamamoto},
        ``Dissipation-Induced Pure Gaussian State,''
        Phys. Rev. A \textbf{85}, 022103 (2012).


     \bibitem{2012-Marcos}{ D Marcos, A Tomadin, S Diehl, and P Rabl},
        ``Photon Condensation in Circuit Quantum Electrodynamics by Engineered Dissipation,''
         New J. Phys. \textbf{14}, 055005 (2012).

     \bibitem{2009-Verstraete}{ Frank Verstraete, Michael M. Wolf, and J. Ignacio Cirac},
        ``Quantum Computation and Quantum-State Engineering Driven by Dissipation,''
         Nature Phys. \textbf{5}, 633 (2009).
     \bibitem{2012-Murch}{ K. W. Murch, U. Vool, D. Zhou, S. J. Weber, S. M. Girvin, and I. Siddiqi},
        ``Cavity-Assisted Quantum Bath Engineering,''
         Phys. Rev. Lett. \textbf{109}, 183602 (2012).

     \bibitem{2006-Pechen}{ Alexander Pechen and Herschel Rabitz},
        ``Teaching the Environment to Control Quantum Systems,''
         Phys. Rev. A \textbf{73}, 062102 (2006).
     \bibitem{2006-Pechen-2}{ Alexander Pechen, Nikolai Il'in, Feng Shuang, and Herschel Rabitz},
        ``Quantum Control by von Neumann Measurements,''
         Phys. Rev. A \textbf{74}, 052102 (2006).

     \bibitem{2013-Kastoryano}{ M. J. Kastoryano, M. M. Wolf, and J. Eisert},
        ``Precisely Timing Dissipative Quantum Information Processing,''
         Phys. Rev. Lett. \textbf{110}, 110501 (2013).

     \bibitem{2011-Barreiro}{J. T. Barreiro, M. Muller, P. Schindler, D. Nigg, T. Monz, M. Chwalla, M. Hennrich, C. F. Roos, P. Zoller, and R. Blatt},
        ``An Open-System Quantum Simulator with Trapped Ions,''
        Nature \textbf{470}, 486 (2011).


     \bibitem{2015-Metelmann}{ A. Metelmann and A. A. Clerk}, 
        ``Nonreciprocal Photon Transmission and Amplification via Reservoir Engineering,''
        Phys. Rev. X \textbf{5}, 021025 (2015).

     \bibitem{2014-Szigeti}{ S. S. Szigeti, A. R. R. Carvalho, J. G. Morley, and M. R. Hush}, 
        ``Ignorance Is Bliss: General and Robust Cancellation of Decoherence via No-Knowledge Quantum Feedback,''
        Phys. Rev. Lett. \textbf{113}, 020407 (2014).

     \bibitem{2013-Lemeshko}{M. Lemeshko and H. Weimer},
        ``Dissipative Binding of Atoms by Non-Conservative Forces,''
        Nat. Commun. \textbf{4}, 2230 (2013).

     \bibitem{2014-Lee}{T. E. Lee and C.-K. Chan},
        ``Heralded Magnetism in Non-Hermitian Atomic Systems,''
        Phys. Rev. X \textbf{4}, 041001 (2014).

     \bibitem{2015-Schempp}{H. Schempp, G. Gunter, S. Wuster, M. Weidemuller, and S. Whitlock},
        ``Correlated Exciton Transport in Rydberg-Dressed-Atom Spin Chains,''
        Physical Review Letters \textbf{115}, 093002 (2015).

     \bibitem{2015-Schonleber}{D. W. Schonleber, A. Eisfeld, M. Genkin, S. Whitlock, and S. Wuster},
        ``Quantum Simulation of Energy Transport with Embedded Rydberg Aggregates,''
        Phys. Rev. Lett. \textbf{114}, 123005 (2015).


     \bibitem{2014-Lesanovsky}{I. Lesanovsky and J. P. Garrahan},
        ``Out-of-Equilibrium Structures in Strongly Interacting Rydberg Gases with Dissipation,''
        Phys. Rev. A \textbf{90}, 011603 (2014).

     \bibitem{2014-Everest}{B. Everest, M. R. Hush, and I. Lesanovsky},
        ``Many-Body out-of-Equilibrium Dynamics of Hard-Core Lattice Bosons with Nonlocal Loss,''
        Phys. Rev. B \textbf{90}, 134306 (2014).


     \bibitem{2014-Woolley}{M. J. Woolley and A. A. Clerk},
        ``Two-Mode Squeezed States in Cavity Optomechanics via Engineering of a Single Reservoir,''
        Phys. Rev. A \textbf{89}, 063805 (2014).

     \bibitem{2012-Tomadin}{ A. Tomadin, S. Diehl, M. D. Lukin, P. Rabl, and P. Zoller},
        ``Reservoir Engineering and Dynamical Phase Transitions in Optomechanical Arrays,''
         Phys. Rev. A \textbf{86}, 033821 (2012).

     \bibitem{2013-Shankar}{S. Shankar, M. Hatridge, Z. Leghtas, K. M. Sliwa, A. Narla, U. Vool, S. M. Girvin, L. Frunzio, M. Mirrahimi, and M. H. Devoret},
        ``Autonomously Stabilized Entanglement between Two Superconducting Quantum Bits,''
        Nature \textbf{504}, 419 (2013).
     \bibitem{2014-Cohen}{J. Cohen and M. Mirrahimi},
        ``Dissipation-Induced Continuous Quantum Error Correction for Superconducting Circuits,''
        Phys. Rev. A \textbf{90}, 062344 (2014).
     \bibitem{2013-Leghtas}{Z. Leghtas, U. Vool, S. Shankar, M. Hatridge, S. M. Girvin, M. H. Devoret, and M. Mirrahimi},
        ``Stabilizing a Bell State of Two Superconducting Qubits by Dissipation Engineering,''
        Phys. Rev. A \textbf{88}, 023849 (2013).


     \bibitem{2016-Bondar}{ D. I. Bondar, R. Cabrera, A. Campos, S. Mukamel, and H. A. Rabitz}, 
        ``Wigner-Lindblad Equations for Quantum Friction,''
        The Journal of Physical Chemistry Letters \textbf{7}, 1632 (2016).


     \bibitem{2005-Petruccione}{F. Petruccione and B. Vacchini},
        ``Quantum Description of Einstein's Brownian Motion,''
        Phys. Rev. E \textbf{71}, 046134 (2005).

     \bibitem{2005-Vacchini}{B. Vacchini},
        ``Master-Equations for the Study of Decoherence,''
        Int. J. Theor. Phys. \textbf{44}, 1011 (2005).

     \bibitem{2009-Vacchini}{B. Vacchini and K. Hornberger},
        ``Quantum Linear Boltzmann Equation,''
        Phys. Rep. \textbf{478}, 71 (2009).



     \bibitem{1995-Holevo}{A. S. Holevo},
        ``On Translation-Covariant Quantum Markov Equations,''
        Izv. Math. \textbf{59}, 427 (1995).

     \bibitem{1996-Holevo}{A. S. Holevo},
        ``Covariant Quantum Markovian Evolutions,''
        J. Math. Phys. \textbf{37}, 1812 (1996).


     \bibitem{1949-Bartlett}{ M. S. Bartlett and J. E. Moyal}, 
        ``The Exact Transition Probabilities of Quantum-Mechanical Oscillators Calculated by the Phase-Space Method,''
        Math. Proc. Camb. Philos. Soc. \textbf{45}, 545 (1949).


     \bibitem{1992-Berg-Sorenson}{K. Berg-Sorenson, Y. Castin, E. Bonderup, and K. Molmer},
        ``Momentum Diffusion of Atoms Moving in Laser Fields''
        J. Phys. B: At. Mol. Opt. Phys. \textbf{25}, 4195 (1992).

     \bibitem{1996-Poyatos}{J. F. Poyatos, J. I. Cirac, and P. Zoller},
        ``Quantum Reservoir Engineering with Laser Cooled Trapped Ions,''
        Phys. Rev. Lett. \textbf{77}, 4728 (1996).

     
     \bibitem{2016-Vuglar}{S. L. Vuglar, D. V. Zhdanov, R. Cabrera, T. Seideman, C. Jarzynski, H. A. Rabitz, and D. I. Bondar},
        ``Quantum Statistical Forces via Reservoir Engineering,''
        arXiv:1611.02736 (2016).



     \bibitem{2009-Brif}{ C. Brif, R. Chakrabarti, and H. Rabitz}, 
        ``Control of Quantum Phenomena: Past, Present, and Future,''
        New J. Phys. \textbf{12}, 075008 (2009).



     \bibitem{2016-Bondar-a}{ D. I. Bondar, R. Cabrera, A. Campos, S. Mukamel, and H. A. Rabitz}, 
        Errata in ``Wigner-Lindblad Equations for Quantum Friction,'' to be published.

     \bibitem{1998-Gao}{S. Gao} 
        ``Dissipative Quantum Dynamics with a Lindblad Functional,''
        Phys. Rev. Lett. \textbf{79}, 3101 (1997).

     \bibitem{1998-Wiseman}{H. M. Wiseman and W. J. Munro}
        ``Comment on `Dissipative Quantum Dynamics with a Lindblad Functional'\thinspace,'' 
        Phys. Rev. Lett. \textbf{80}, 5702 (1998).

     \bibitem{2000-Vacchini}{B. Vacchini},
        ``Completely Positive Quantum Dissipation,''
        Phys. Rev. Lett. \textbf{84}, 1374 (2000).
     \bibitem{2001-O'Connell}{R. F. O'Connell},
        ``Comment on `Completely Positive Quantum Dissipation'\thinspace,''
        Phys. Rev. Lett. \textbf{87}, 028901 (2001).
     \bibitem{2001-Vacchini}{B. Vacchini},
        ``Vacchini Replies,''
        Phys. Rev. Lett. \textbf{87}, 028902 (2001).

     \APPBIBITEM
\end{thebibliography}
\end{document}